\documentstyle[emulateapj]{article}

\lefthead{von Braun and Mateo}
\righthead{NGC 3201 Reddening Map and CMD}

\begin{document}

\title{An Extinction Map and Color Magnitude Diagram for the Globular
Cluster NGC 3201}

\author{Kaspar von Braun and Mario Mateo}

\affil{University of Michigan}
\authoraddr{Department of Astronomy, Univ. of Michigan, Ann Arbor, MI 48109-1090}
\authoremail{kaspar@astro.lsa.umich.edu, mateo@astro.lsa.umich.edu}

%--------------------------------------------------------------------------
\begin{abstract}
%--------------------------------------------------------------------------

Differential $E_{V-I}$ variations of up to $\sim 0.2$ mag on a scale
of arcminutes across NGC 3201 are presented in the form of an
extinction map. This map, created by calculating average $E_{V-I}$
values for stars in small subregions of the field with respect to a
fiducial region, greatly improves the appearance of the CMD of the
cluster.  We describe how we implemented this technique in detail with
our data for NGC 3201.  A comparison between our map and that of the
same region extracted from the COBE/DIRBE reddening maps published by
Schlegel, Finkbeiner, \& Davis (1998)\markcite{schlegel98} (SFD)
displays larger-scale similarities between the two maps as well as
smaller-scale features which show up in our map but not in the SFD
map.  Several methods of determining an $E_{V-I}$ zeropoint to add to
our differential extinction map are presented. Isochrone fitting
proved to be the most successful one, but it produces an average
$E_{V-I}$ for the cluster which is smaller than previously published
values by $\sim 1.5\sigma$.  Finally, our results seem to support the
statement by Arce \& Goodman (1999)\markcite{arce99} that the SFD maps
overestimate the reddening in regions of high extinction.

%--------------------------------------------------------------------------
\end{abstract}
%--------------------------------------------------------------------------

%--------------------------------------------------------------------------
%--------------------------------------------------------------------------

%--------------------------------------------------------------------------
\section{Introduction}
%--------------------------------------------------------------------------

We are currently undertaking a survey of approximately 15 Galactic
globular clusters (GCs) with the aim of identifying eclipsing binary
stars around or below the turn-off-point by means of detecting
brightness variations.  These systems can provide masses for
Population II main sequence stars. High-quality photometry is
essential for obtaining reliable values for surface temperatures,
luminosities, ages, etc.  Precise extinction determinations are a
critical part of obtaining these results from our photometry.

NGC 3201, by chance one of the first GCs we analyzed, is a
low-latitude cluster located at $\alpha_{2000} = 10^{h} 36^{m}
36.8^{s}$ and $\delta_{2000} = -46^{\circ} 24^{'} 40^{''}$ or $l =
277.2^{\circ}$ and $b = 8.6^{\circ}$ (Harris 1996)\markcite{h96} with
a retrograde orbit around the center of the Milky Way.  Its nearby
location ($d$ = 5.2 kpc) and low concentration ($c$ = 1.31) (Harris
1996) make it an attractive target for photometric studies.  Recent
such studies include Cacciari (1984),\markcite{cacciari84} Brewer et
al. (1993),\markcite{brewer93} C\^{o}t\'{e} et
al. (1994),\markcite{cote94} Covino \& Ortolani
(1997)\markcite{covino97} (hereafter C\&O), and Gonzalez \&
Wallerstein (1998)\markcite{gonzalez98} (hereafter GW).

Due to its low-latitude position, the effects of differential
reddening across the face of NGC 3201 are quite substantial.  The
existence of variable extinction was noted in practically all earlier
studies of this cluster.  Cacciari (1984), for instance, finds an
irregular reddening distribution with a variation of $\Delta E_{B-V}$
= 0.03 mag in addition to a mean value of $E_{B-V}$ = 0.21. GW report
a range in $E_{B-V}$ of as much as 0.1 mag across NGC 3201.  Using
spectroscopic data for 18 giants in NGC 3201, GW correct for the color
spread by modelling the reddening in $E_{B-V}$ as a plane in $\alpha$
and $\delta$.

Recently published dust infrared emission maps by Schlegel,
Finkbeiner, \& Davis (1998)\markcite{schlegel98} (SFD hereafter) seem
to indicate, however, that the dust distribution in the region of NGC
3201 is too clumpy to be fitted by a linear function in position, even
on the scale of arcminutes (see Fig. ~\ref{schlegelmap}).

In this work, we attempt to remove the differential reddening across
the cluster by using our high-quality $VI$ photometry. Our procedure
is, in principle, similar to the one used by Piotto et
al. (1999).\markcite {piotto99} It aims to find the average $\Delta
E_{V-I}$ for subregions of the cluster field of approximately 1
arcmin$^{2}$ in size with respect to a fiducial region in the cluster
where little or no differential reddening is apparent and where the
overall $E_{V-I}$ is small compared to the rest of the cluster.  The
pixel size (resolution) of this extinction map is approximately one
fourth of the size (in area) of the SFD maps.  Since the SFD maps are
based on infrared emission from dust, our map will provide an
independent check of whether, if at all, they tend to overestimate
reddening, as was suggested by, e.g., Arce \& Goodman
(1999)\markcite{arce99} for regions with $A_{V} > 0.5$ mag.

Our results indicate variations in $E_{V-I}$ of up to $\sim$ 0.2 mag.
Differential reddening this strong can wreak havoc with photometric
and spectroscopic studies of cluster stars. The inherent uncertainties
in the various parameters of the CMD (e.g., magnitude and color of the
main-sequence turnoff (MSTO), horizontal branch features, etc.)  are
greatly amplified. Moreover, one may ``detect'' age or
surface-temperature gradients where there are none in reality. To give
an example of the magnitude of the effect, using the recently
published color-temperature relations by Houdashelt, Bell, \& Sweigart
(2000),\markcite{houdashelt00} the effective temperature of a
solar-metallicity, main sequence star with $V-I \sim 0.8$ would vary
by approximately 600$K$ for a differential reddening effect of
$E_{V-I} \sim 0.2$ mag, which, in turn, could lead to errors in the
metallicity determination.

Using our internal dereddening technique, we obtain a high-quality,
deep-photometry CMD of NGC 3201 comprised of approximately ninety 600s
exposures as well as some shorter ones, all using the $V$ and $I$
bands.

Our observations and data reductions, as well as the details about our
internal dereddening method, are documented in Section 2.  Section 3
contains our results concerning the reddening map and the CMD of NGC
3201.  Finally, we discuss the determination of the reddening
zeropoint of our extinction map in Section 4 and give a brief summary
of our work in Section 5.

%--------------------------------------------------------------------------
\section{Observations, Data Reduction, and Internal Dereddening}
%-------------------------------------------------------------------------

%--------------------------------------------------------------------------
\subsection{Observations and Basic Data Reductions}
%--------------------------------------------------------------------------

The NGC 3201 observations were obtained during the nights of April 26
through May 6, 1998, at the Las-Campanas-Observatory (LCO) 1m Swope
Telescope, using Johnson-Cousins $VI$ filters and a SITe 1 $2048^{2}$
CCD with a field-of-view of 23.5 arcmin on a side.  Table 1 gives the
number of epochs we observed for different exposure times which were
chosen in order to cover a larger magnitude range in the CMD.

\placetable{table1} 

The initial processing of the raw CCD images was done with the
routines in the IRAF\footnote{IRAF is distributed by the National
Optical Astronomy Observatories, which are operated by the Association
of Universities for Research in Astronomy, Inc., under cooperative
agreement with the NSF.} CCDPROC package. For each night, 10 bias
frames were combined for the bias subtraction.  The $V$ band flats
were produced by combining between 4 and 6 twilight flat images per
night. All $I$ band data were first flattened using an image comprised
of 15 individual $I$ band domeflats, and then divided by a normalized,
dome-flattened dark-sky-flat which itself was created by
median-averaging approximately 40 individual different blank sky
fields. Finally, the $I$ band images were corrected for fringing by
subtracting a fringe image which was created by subtracting the mean
pixel value of the dark-sky-flat from the dark-sky-flat image itself.
 
The processed data were reduced using DoPHOT (Schechter, Mateo, and
Saha 1993)\markcite{schech93} in ``fixed-position mode'' where the
positions of stars are fixed (after correcting for global
frame-to-frame shifts as well as small distortions) to the positions
measured on a deep-photometry template image obtained by co-adding the
$\sim$ 15 best-seeing frames.

The photometric results for every star were averaged over all frames
with the same exposure time to obtain the final magnitude for the star
under investigation. For the 600s exposures, our only requirement was
that a star appear in more than 75\% of the epochs. The 60s images
were taken to complete the CMD in the brighter regions; we only took
magnitudes from these exposures of stars which were saturated in the
600s exposures. The same procedure was followed for the 10s frames.
Aperture corrections were applied by calculating one constant value
for the whole chip after we found little dependence upon position on
the CCD.  The data from the shorter exposures were shifted to the
photometric system of the 600s exposures using non-saturated stars in
common.

During our consecutive two photometric nights (May 2 \& 3, 1998), we
obtained a total of more than 140 observations of various Landolt
(1992)\markcite{land92} standard stars which covered a range of $-0.2
< V-I < 2.2$ in color as well as $1.15 < X < 1.55$ in airmass.  Using
the IRAF PHOTCAL package, we applied one single standard star solution
for data from both nights of the form

\begin{equation}
 V = v + a_{0} + b_{0} X_{v} + c_{0} (v-i)
\end{equation}
\begin{equation}
 V - I = a_{1} + b_{1} X_{i} + c_{1} (v-i),
\end{equation}

\noindent
where the $a_{j}$, $b_{j}$, and $c_{j}$ are the fitted constants,
$X_{filter}$ is the airmass of the exposure taken with the respective
filter, the lowercase magnitudes are instrumental, and the uppercase
ones are the known magnitudes of the standard stars.  The root mean
square error was 0.007 mag and 0.016 mag for $V$ and $V-I$,
respectively

Astrometry was performed by identifying 59 USNO reference stars (Monet
et al. 1996)\markcite{monet96} in NGC 3201 and using the IRAF IMAGES
package for the coordinate transformation. A linear fit in $x$ and $y$
produced errors around 0.2 arcsec, consistent with the USNO
precision. The tangent point of the transformation was $(x,y) =
(962.7,934.2)$ and corresponds to $\alpha_{2000} = 10^{h} 17^{m}
31.2^{s}$ and $\delta_{2000} = 46^{\circ} 24^{'} 11^{''}$. The
rotation of $x$ and $y$ with respect to $\alpha$ and $\delta$ was
$91.23^{\circ}$, and the pixel scale came out to be 0.695
pix/arcsec. These results enabled us to create the reference grid in
Fig. \ref{extmap}.

%--------------------------------------------------------------------------
\subsection{Mapping the Differential Reddening}
%--------------------------------------------------------------------------

\begin{figure}
\plotone{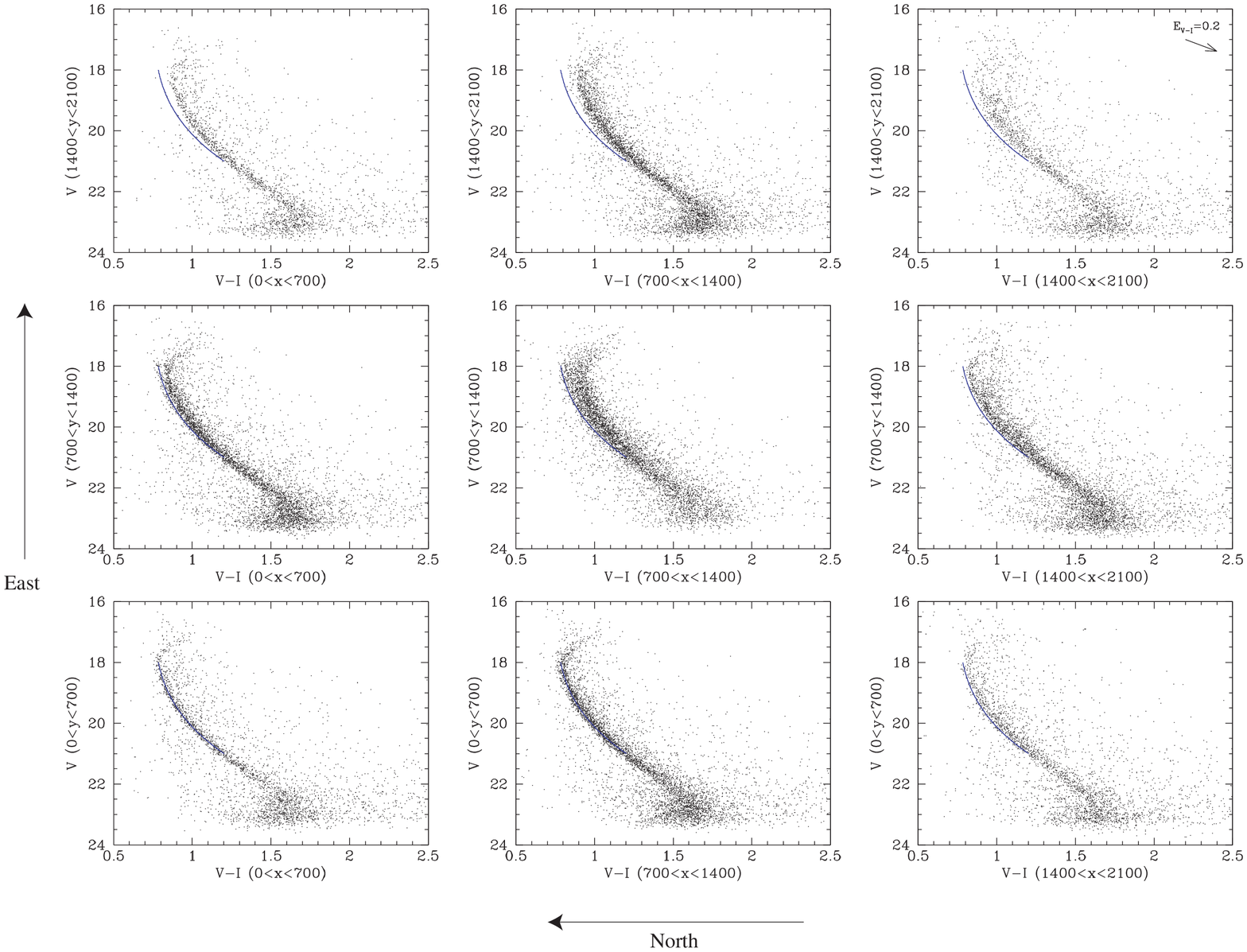}
\caption{``Raw'' (calibrated but before dereddening) CMDs of 9
subregions of the field of view of NGC 3201.  The $x$ and $y$ ranges
of the respective subregion are given on the axes of each individual
CMD.  $x = 0, 700, 1400, 2100$ approximately correspond to
$\delta_{2000} = -46^{\circ}13^{'}03^{''}, -46^{\circ}21^{'}09^{''},
-46^{\circ}29^{'}16^{''}$, and $-46^{\circ}37^{'}22^{''}$,
respectively; $y = 0, 700, 1400, 2100$ approximately correspond to
$\alpha_{2000} = 10^{h}16^{m}27^{s}, 10^{h}17^{m}14^{s},
10^{h}18^{m}01^{s}$, and $10^{h}18^{m}48^{s}$, respectively.  The
directional arrows on the bottom and the left of the image show the
approximate orientation of the CCD.  The varying appearance and
broadness of the main sequence as a function of position clearly
indicates the differential reddening across the field of view. The
line through the CMDs represents the best fit through the datapoints
in the fiducial region $1200<x<1600$ and $0<y<400$ (described in
Section 2.2, item 1).  It is apparent that there is not only less
differential reddening in that region (tight main sequence), but also
that the overall $E_{V-I}$ is low. The reddening vector for
$E_{V-I}=0.2$ is shown in the top right panel.
}\label{raw_cmd}
\end{figure}

\begin{figure}
\plotone{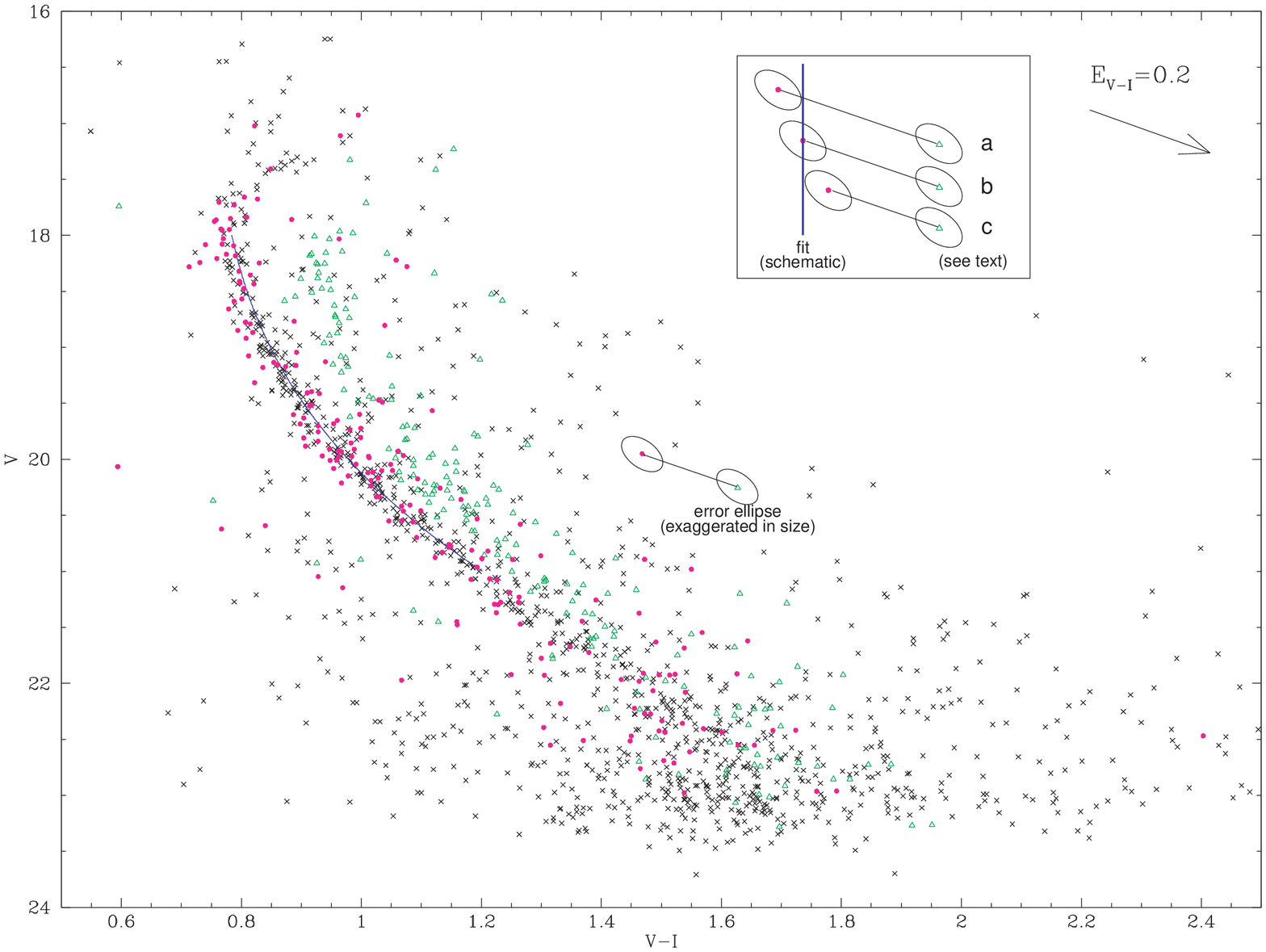}
\caption{This figure illustrates the dereddening procedure which was
used in this paper. The triangles represent stars in the subregion
$700 < x < 800$ and $800 < y < 900$ (approximately $10^{h}17^{m}21^{s}
< \alpha_{2000} < 10^{h}17^{m}28^{s}$, as well as
$-46^{\circ}21^{'}07^{''} < \delta_{2000} < -46^{\circ}22^{'}16^{''}$)
before being dereddened; the filled circles represent the same stars
after the dereddening along the reddening vector shown in the top
right corner of the figure. The x-shaped symbols are the stars in the
fiducial region (mentioned above) in which none or very little
differential reddening is taking place. The fitted polynomial with
range $18<$ mag $<21$ is visible as the curve through the fiducial
region datapoints.  The reddening for this region is close to
$E_{V-I}=0.2$ with respect to the average of the fiducial region.  The
tilted error ellipses, described in Section 2.2, item 7, greatly
exaggerated in size, are shown to illustrate our dereddening method
which is schematically outlined in the box to the upper right. The
straight line represents the fit through the datapoints in the
fiducial region.  Example {\bf a} represents the ``point of last
contact'' of the error ellipse with the fit, example {\bf b} the shift
from the original datapoint until it intersects the fit, and example
{\bf c} the ``point of first contact'' between the error ellipse and
the fit.\label{procedure}}
\end{figure}

At this point in the data reduction we were in the possession of a
fully calibrated CMD whose main sequence appeared exceptionally broad
(see Fig. \ref{cmd_raw}). After ruling out a number of possible
instrumental causes for this effect, and after we noted a clear
dependence of the appearance of the main sequence upon the positions
of stars on the CCD (illustrated in Fig. \ref{raw_cmd}), we concluded
that the underlying cause for the broadness of our main sequence was
differential extinction across the field of view. This seemed
especially sensible given the low-latitude location of NGC 3201 and
motivated us to create a reddening map since our binary star data were
not useful without this correction.  In order to create such a map for
NGC 3201, we developed the steps described below (see also
Fig. \ref{procedure}):

\placefigure{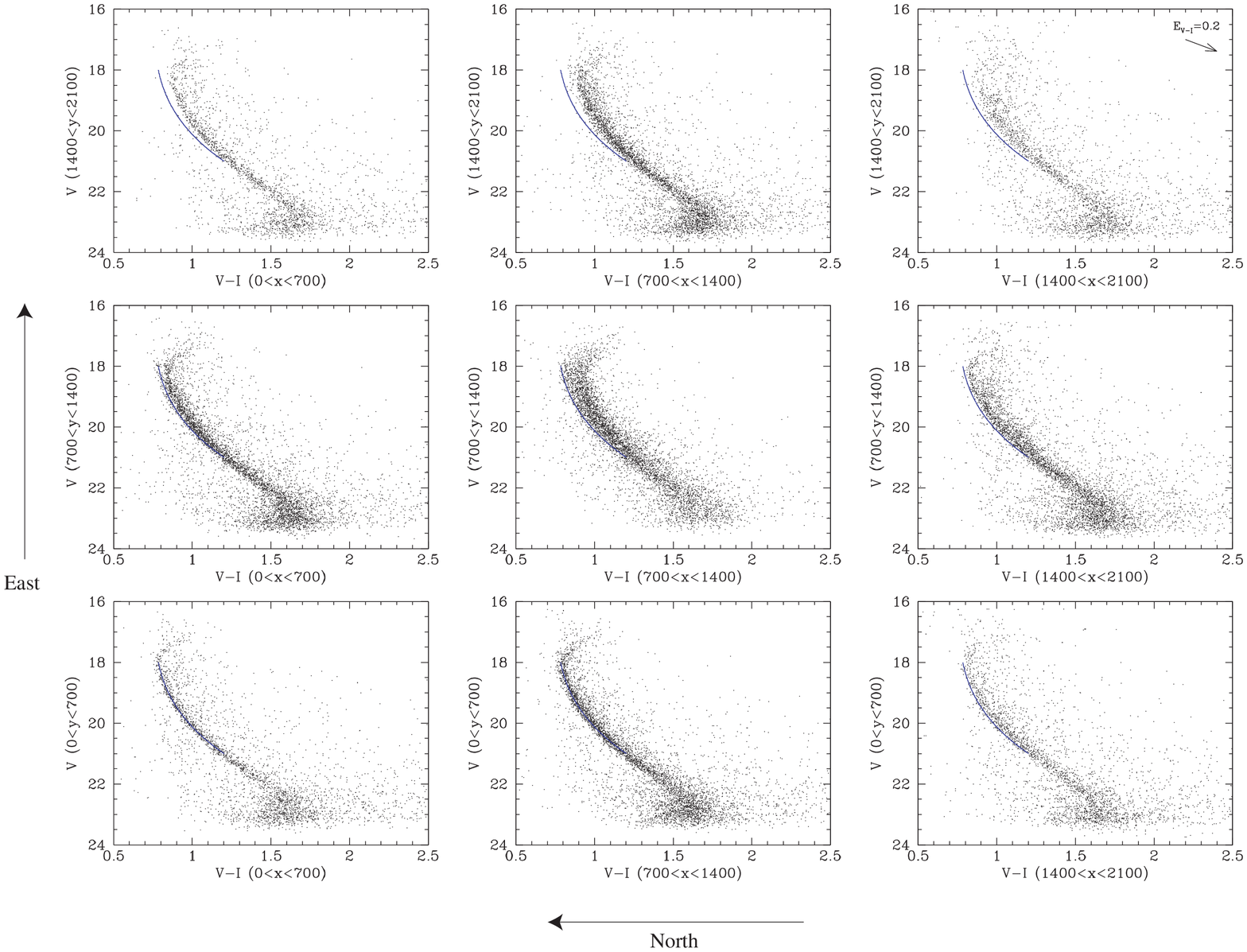}

\placefigure{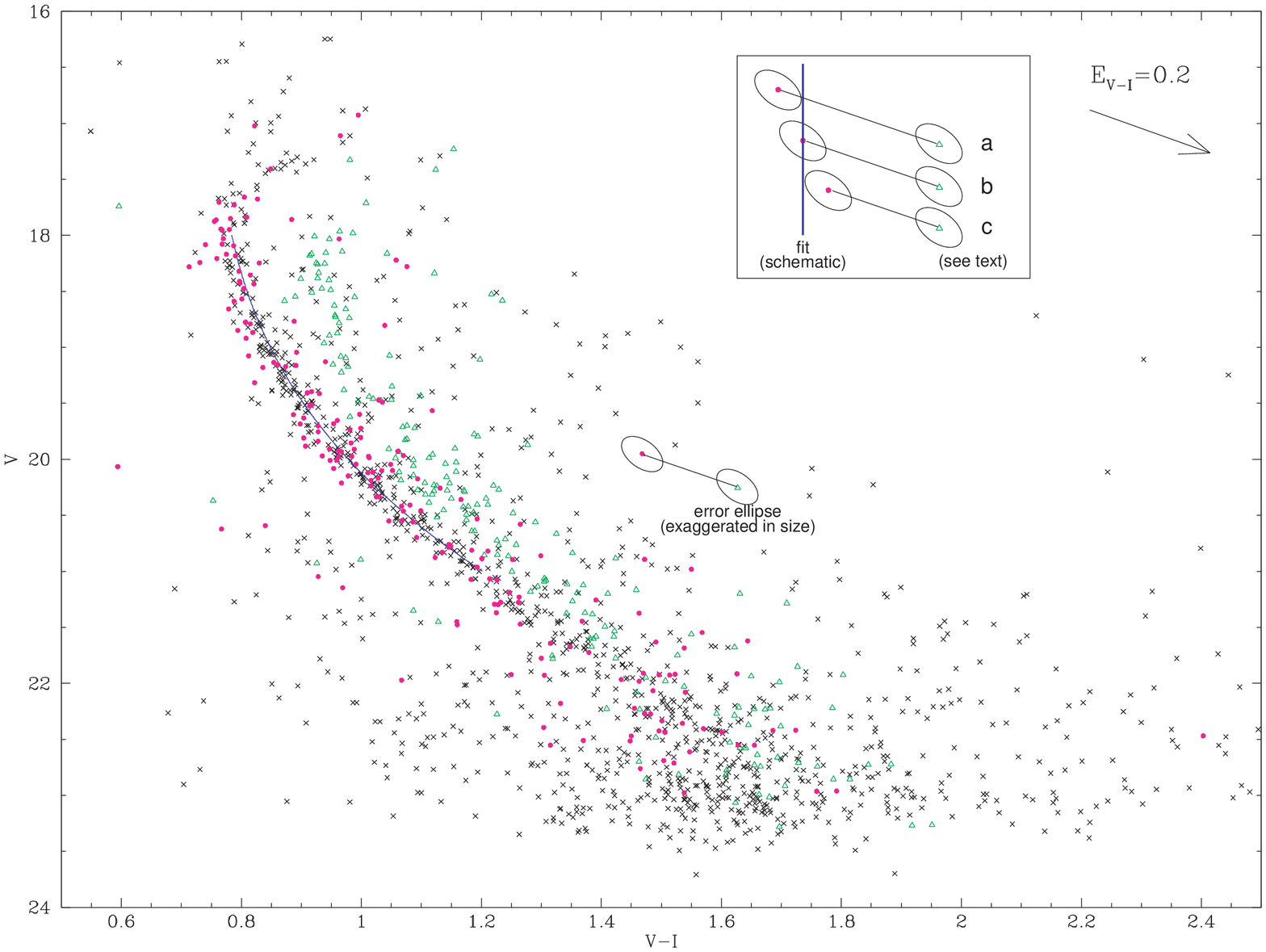}

\begin{enumerate}

\item A fiducial region was chosen in which 
\begin{itemize}
\item very little or no differential 
reddening occurred, i.e., where the main sequence appeared narrow;
\item the overall $E_{V-I}$ was very small with respect to 
the rest of the field of view;
\item there were enough stars in the field to obtain a statistically 
reliable fit to their positions on the CMD. 
\end{itemize}

The coordinates of this fiducial region are approximately
$10^{h}16^{m}26^{s} < \alpha_{2000} < 10^{h}16^{m}53^{s}$ and
$-46^{\circ}31^{'}21^{''} < \delta_{2000} < -46^{\circ}26^{'}44^{''}$
(see Figures \ref{extmap} through \ref{data-map}).

\item A high-order polynomial was fit to the main sequence stars with
$18 < V < 21$ and $0.7 < V-I < 1.4$ in the fiducial region, using
algorithms supplied by Robbie Dohm-Palmer (2000, private
communication) which enabled us to separate the main sequence stars
from background stars in the CMD, and by Press et
al. (1992)\markcite{press92}, which provided the least-squares fit to
the data (see Fig. \ref{procedure}).

\item We divided the field of view into subregions of different sizes
in order to have a sufficient number of stars in each region.  Towards
the outer parts of the field, the stellar density decreases.  As a
result, the sizes of the subregions increase. The inner subregions are
all of the same size (100$\times$100 pixel). See Fig. \ref{extmap}.

\item For each of these regions, every star falling between $17.9 < V
< 21.1$ and $0.65 < V-I < 1.45$ was incrementally moved along the
reddening vector defined by the relations in Cardelli, Clayton, \&
Mathis (1989)\markcite{card89} until it intersected the fit (see
Fig. \ref{procedure}).

\item The statistical biweight (see Beers, Flynn, \& Gebhardt
1990)\markcite{beers90} of all these incremental shifts of the stars
in a given subregion was then calculated, outliers (0.9 $\sigma$) were
removed, and it was recalculated. This provided the value of the total
shift along the reddening vector for each star. Since the slope of the
reddening vector for a standard extinction law is known (1.919; see
Cardelli, Clayton, \& Mathis (1989)), the differential $E_{V-I}$ for
each star corresponds to the $(V-I)$-component of the vector described
above (Fig. \ref{procedure}).

\item For six of the 100$\times$100 pixel subregions as well as one
200$\times$300 pixel subregion on the extinction map shown in
Fig. \ref{extmap}, too few stars were present to calculate the average
reddening. For these cases, we calculated $E_{V-I}$ by simply
averaging the reddening values of all the neighboring
subregions\footnote{In this process, every 100$\times$100 subregion
was treated individually, even if it was a part of a larger one. That
is, for the 200$\times$300 region just south of the north-east corner
of the field of view in which the number of stars was too low to
obtain a reddening value, the above procedure was applied six times.}.

\item Our error analysis for the shifts described above involved the 
following:
\begin{itemize}
\item We created an error ellipse for every star defined by the values
of the associated errors (as returned by DoPHOT) in color and
magnitude.
\item Since the color and magnitude errors are correlated, the error
ellipse is tilted, with the tilt angle and the lengths of the
semimajor and semiminor axes functions of $\sigma_{V}$ and
$\sigma_{V-I}$.
\item One average tilt angle was obtained for every subregion under
investigation by averaging the results for the individual stars.  For
every subregion, this angle was approximately
$35^{\circ}\pm4^{\circ}$. The semimajor and semiminor axes for every
subregion were obtained in the same way, so that every star in a given
subregion has the same error ellipse associated with it in the center.
\item This error ellipse was shifted along with its respective star
during the dereddening process described above in items 4 and 5.  The
``point of first contact'' (pfc), i.e, the point at which the error
ellipse first touches the fit through the data in the fiducial region,
as well as the ``point of last contact'' (plc), represent the
respective one-sigma-deviation points (see Fig. \ref{procedure}).
\item These two contact points are not necessarily symmetric about the
reddening value of the star (center of error ellipse), but in order to
increase the readability of this publication, we calculated the mean
pfc and plc for a given subregion, and then their average distance
from the center of the ellipse, so that $\sigma_{ellipse} = 0.5(pfc -
plc)$.
\item The final error estimate for each subregion ($\sigma_{E_{V-I}}$)
given as the lower number in each of the pixels in
Fig. \ref{extmapgrid} is then obtained by adding in quadrature
$\sigma_{ellipse}$ and the error in the mean of the shifts of all the
stars in the subregion under investigation.
\end{itemize}

\end{enumerate}

%--------------------------------------------------------------------------
\section{Results}
%--------------------------------------------------------------------------

%--------------------------------------------------------------------------
\subsection{Extinction Map and Comparison with SFD Map}
%--------------------------------------------------------------------------

\placefigure{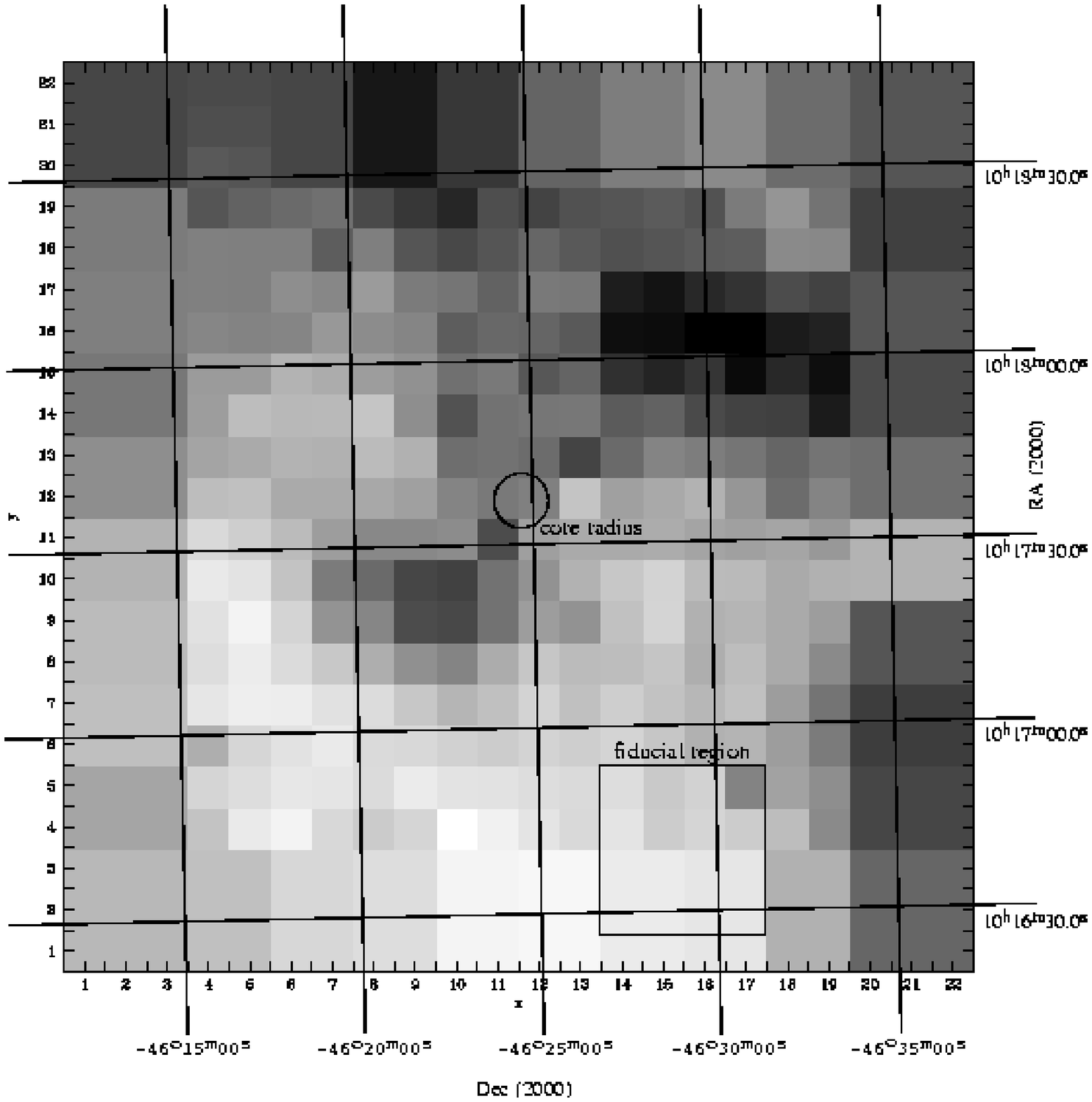}

\placefigure{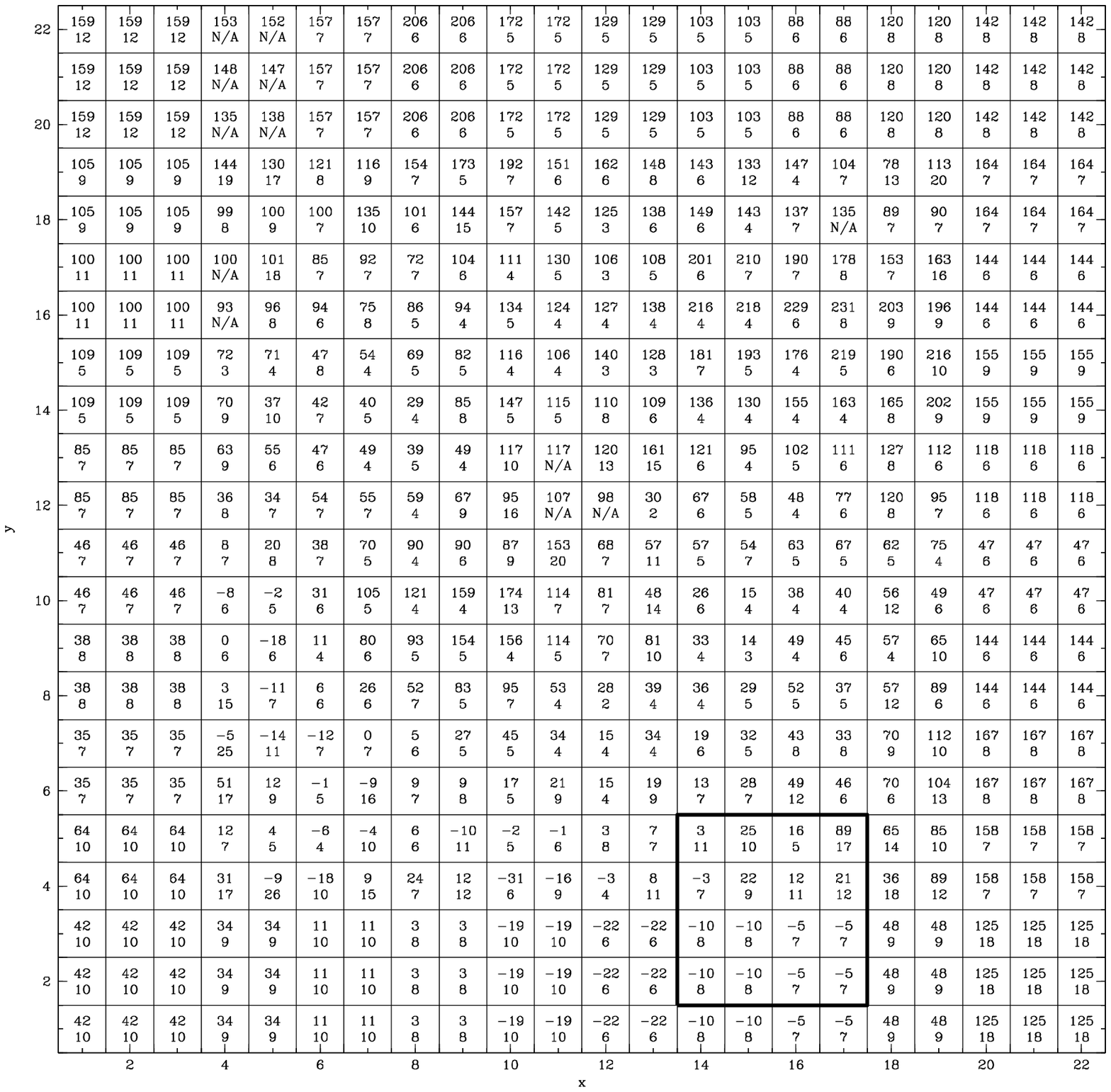}

\placefigure{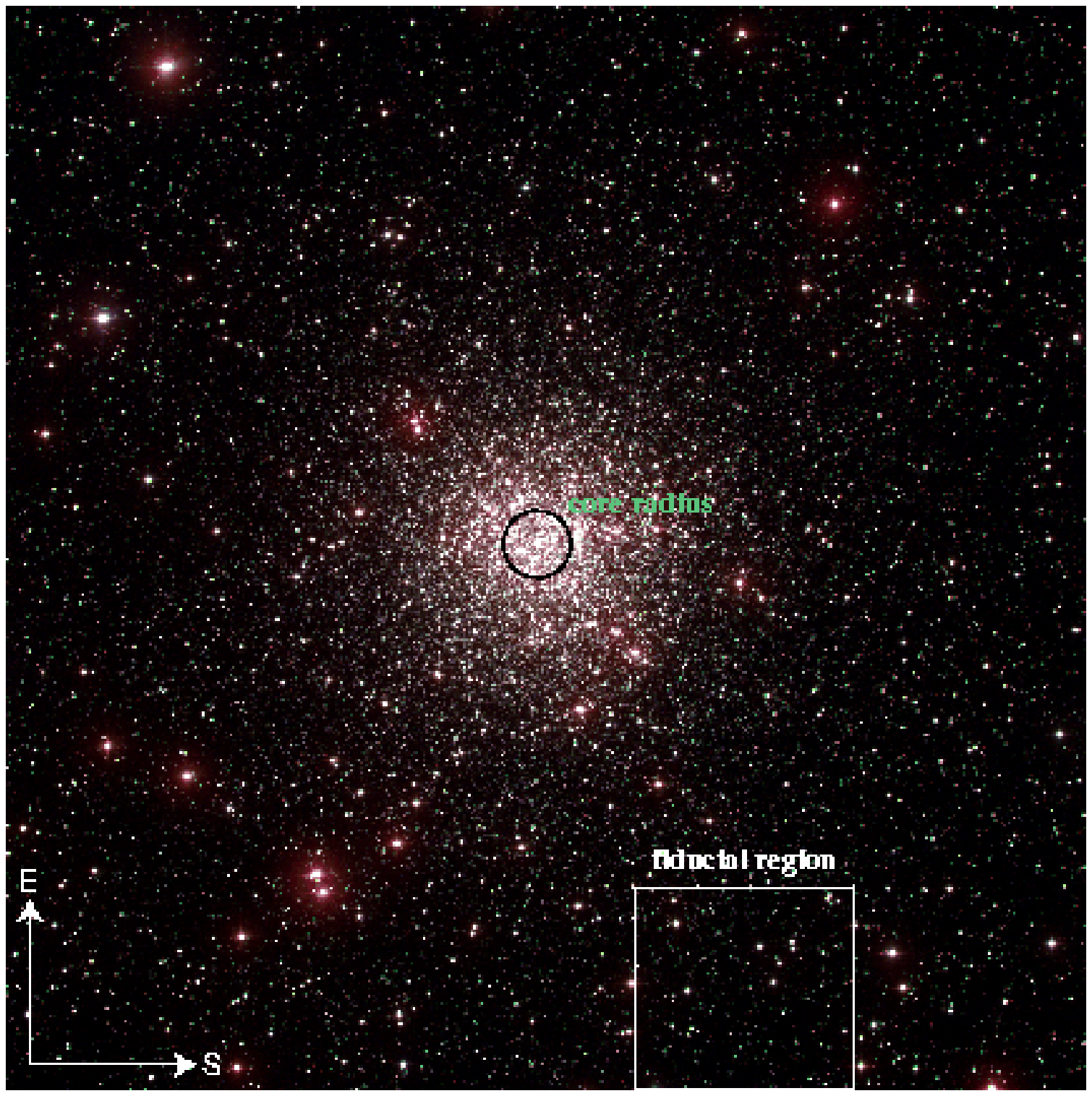}

Fig. \ref{extmap} summarizes the results of our internal dereddening
procedure. Darker regions correspond to higher reddening with respect
to the average reddening taking place in the fiducial region described
in Section 2.2, item 1.  For reference purposes, we included the
approximate locations of this fiducial region and the core radius of
NGC 3201, as well as a coordinate grid.  The average $E_{V-I}$
relative to the fiducial region is 86 mmag with a standard deviation
of 61 mmag.  The individual subregions' $E_{V-I}$ values in
millimagnitudes (relative to the average reddening in the fiducial
region) are shown as the top number in each of the pixels in the grid
of Fig. \ref{extmapgrid}.  The bottom number in each pixel corresponds
to the error in the mean for the corresponding $E_{V-I}$ value. In
addition, we included an image of NGC 3201 (Fig. \ref{ngc3201_final})
in order to indicate the locations of the fiducial region and the core
radius of the globular cluster.

\placefigure{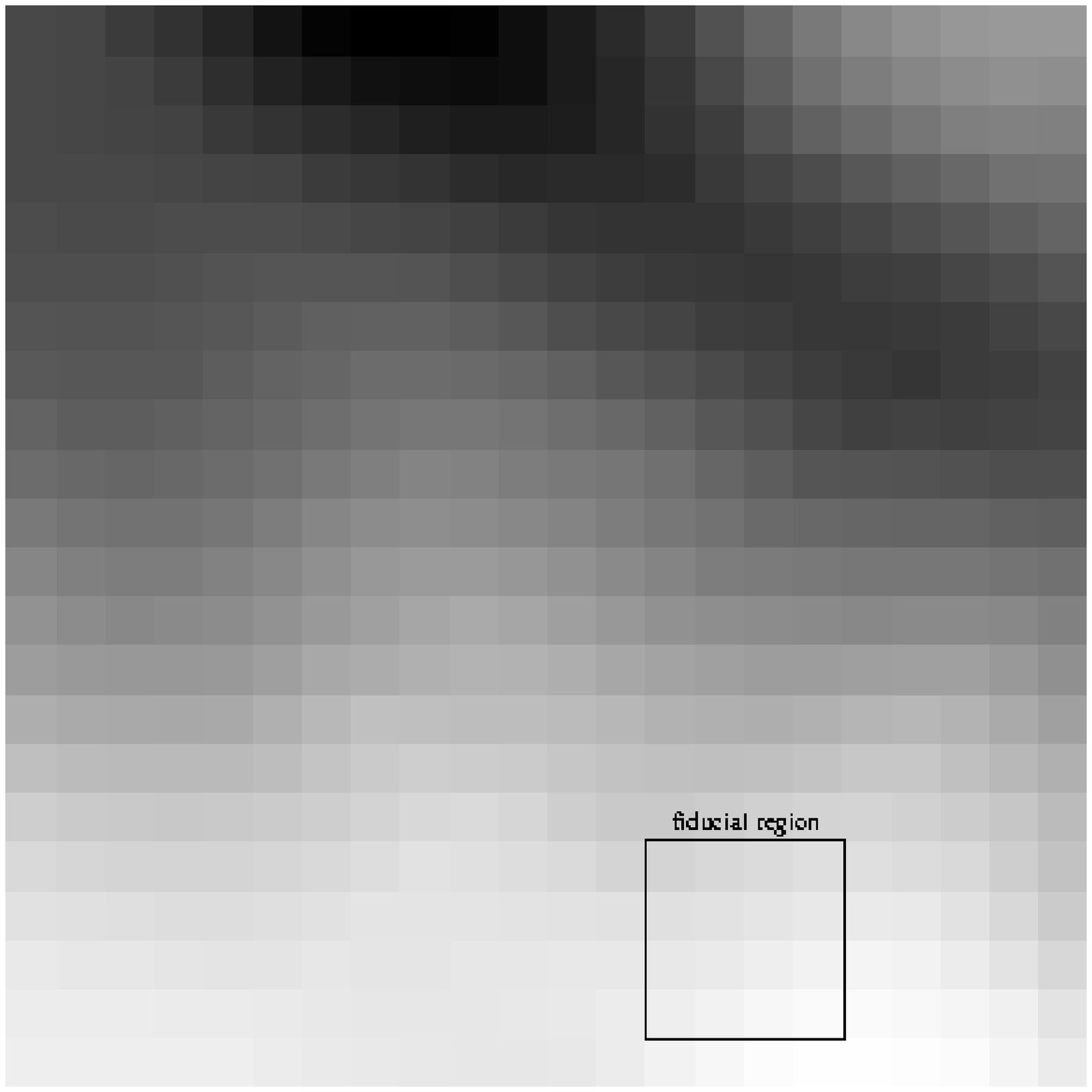}

\placefigure{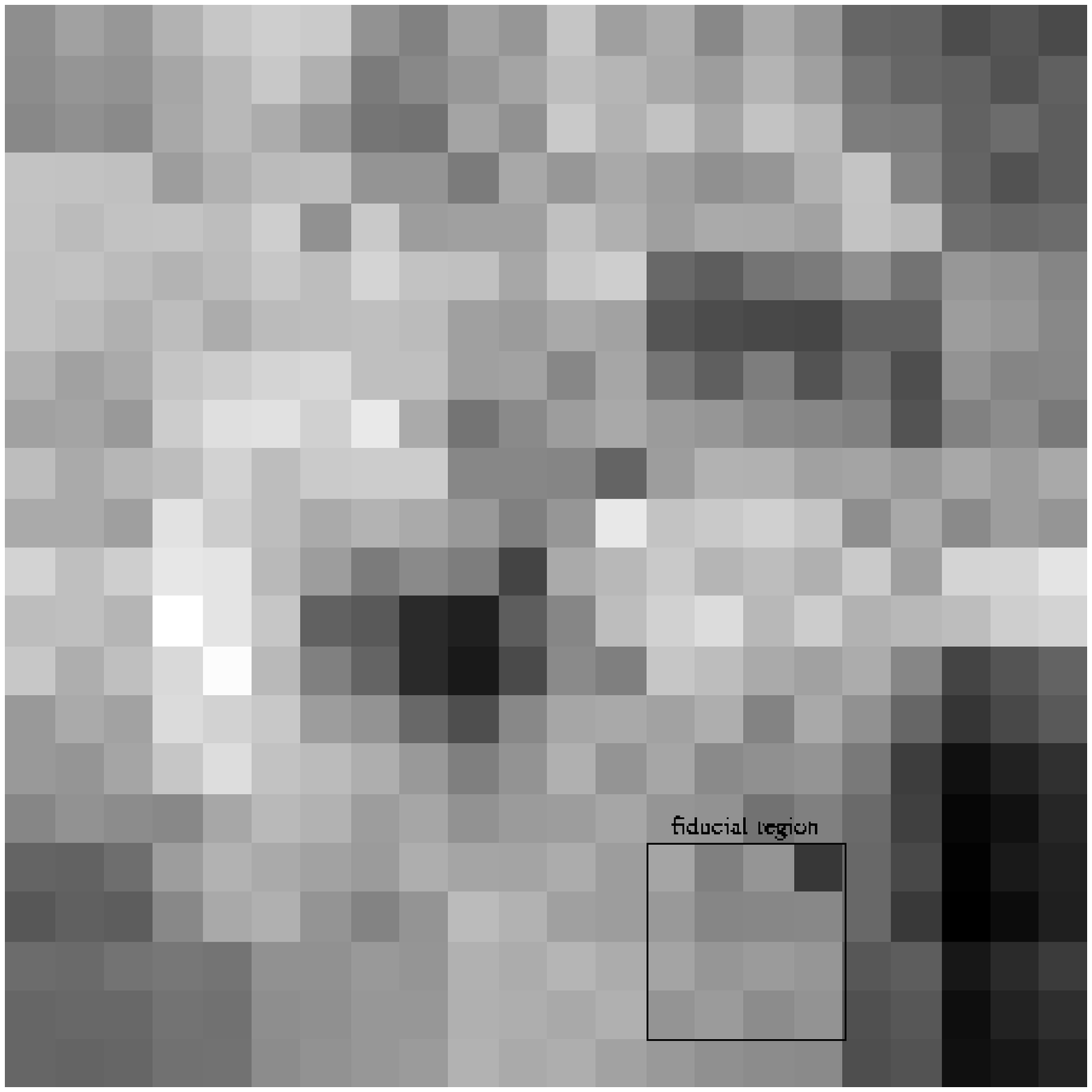}

At this point, these reddening values are all relative to our fiducial
region which itself is suffering some mean interstellar extinction.
The procedure described cannot determine the absolute reddening
zeropoint. Thus, we had to introduce additional information, derived
from previous studies, in order to calculate this zeropoint. The
discussion of this process is given in Section 4.1.

As a first step, we subtracted our extinction map from the SFD map of
the region of NGC 3201 which we observed.  The SFD map of the cluster
region is shown in Fig. \ref{schlegelmap} with the same orientation
and size as Fig. \ref{extmap}. Once again, dark corresponds to higher
extinction.  We added the location of the fiducial region for
reference. The average $E_{V-I}$ of the SFD map is 487 mmag with a
standard deviation of 59 mmag.

A first comparison indicates that, while the larger-scale structure is
the same for both extinction maps (in particular the obscuration ridge
extending from the top left part of the image to the right center part
of it), our map shows smaller-scale features which are not present in
the SFD maps. These features are visible in the difference image in
Fig. \ref{data-map} whose orientation and size is once again identical
to the other Figures. Darker regions correspond to places where our
data indicate the presence of dust which does not show up in the SFD
map; lighter regions show good agreement between the two maps. The
average $E_{V-I}$ of this difference map is 401 mmag with a standard
deviation of 46 mmag (significantly smaller than the standard
deviation of both our extinction map and the SFD map).  This lower
standard deviation is an indication that both our map and the SFD map
of the region of NGC 3201 are tracing many of the same features in the
foreground extinction along the line of sight to the globular cluster.

The smaller-scale features present in only our map but absent in the
SFD map are certainly worth further attention. At this point, we are
not entirely confident that we can give a definite reason for the
discrepancies, but some potential reasons could be the following:

\begin{itemize}

\item The smaller-scale features which are not in the SFD maps might 
have been smoothed out by the lower resolution of the IR observations
used to create the maps.

\item The temperature of the dust producing the smaller-scale features
was outside the detection range of the instrument used to collect the 
data for the SFD maps. This way, they would still show up in our analysis.

\item The dust we are seeing is actually part of the globular cluster
itself\footnote{The phenomenon of dust inside globular clusters was
discussed by, e.g., Forte et al. (1992).\markcite{forte92}} and was
therefore not picked up in the SFD maps.

\end{itemize}

%--------------------------------------------------------------------------
\subsection{Photometry and CMD}
%--------------------------------------------------------------------------

\placefigure{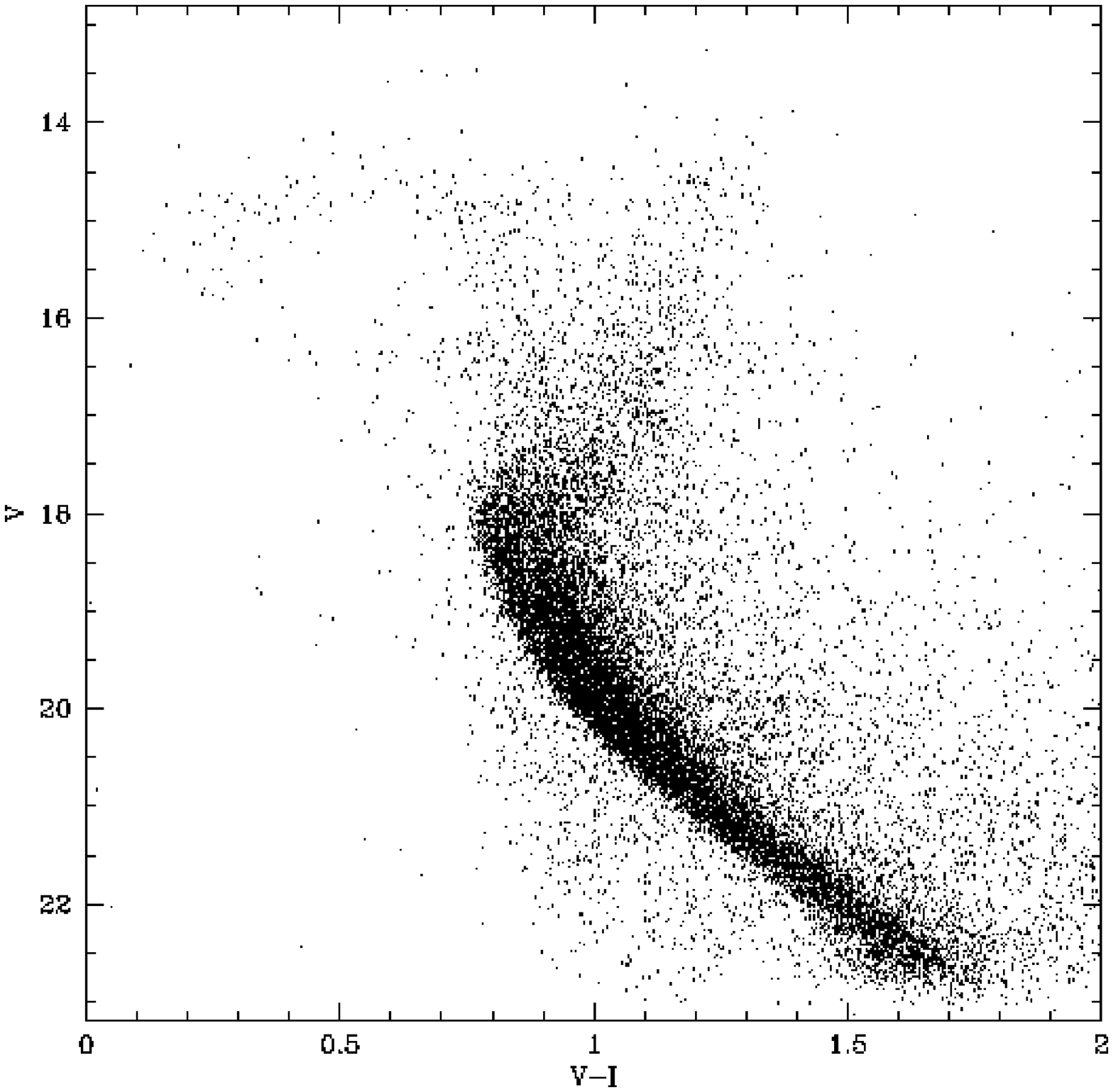}

\placefigure{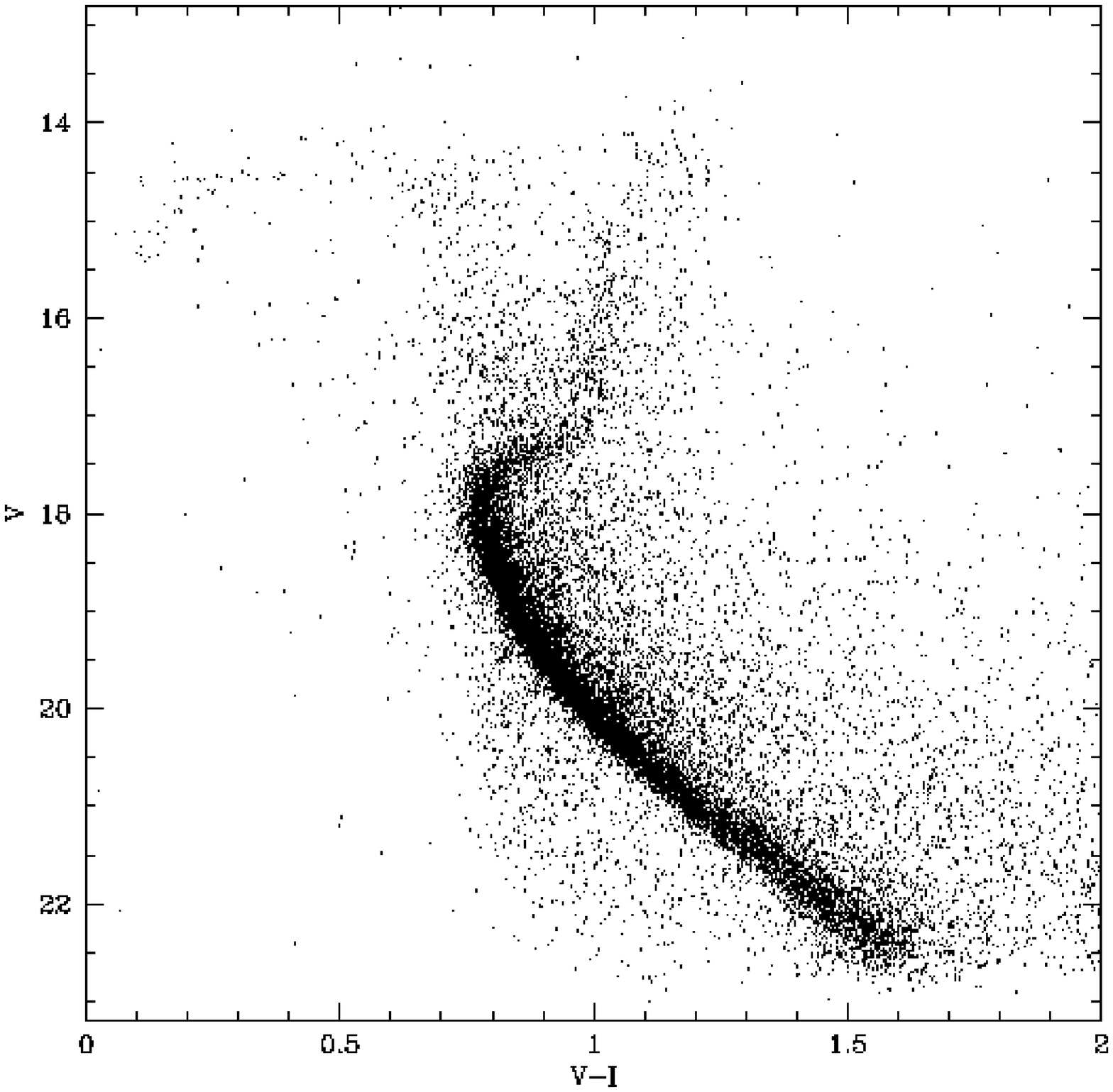}

Fig. \ref{cmd_raw} contains the raw CMD (no dereddening applied) for
NGC 3201. To illustrate the effects of our dereddening method, we
show, in Fig. \ref{cmd_final}, the CMD after applying the extinction
map (differential reddening correction) to the data. The datapoints
are now all shifted to the CMD-location of the stars in the fiducial
region, that is, no absolute reddening zeropoint has been applied. The
improvement is quite stunning. The width of the main sequence of
Fig. \ref{cmd_final} has, by applying the differential reddening map,
decreased to a fraction of its former value in Fig. \ref{cmd_raw}. The
subgiant branch which was hardly visible at all in the raw CMD is now
well traced out by the datapoints. Even the horizontal branch, though
sparsely populated in both CMDs due to saturation effects of the
bright stars in the field, is narrower in the differentially
dereddened CMD.

%--------------------------------------------------------------------------
\section{The Reddening Zeropoint of our Extinction Map for NGC 3201}
%--------------------------------------------------------------------------

%--------------------------------------------------------------------------
\subsection{Comparing the MSTO Color with Previous Studies}
%--------------------------------------------------------------------------

In order to make our reddening map (Fig. \ref{extmapgrid}) a useful
tool, we need to determine the reddening zeropoint to add to the
$E_{V-I}$ values in the grid. As mentioned above, this could not be
done using our data alone. In order to be able to use results from
previous studies for the calculation of this zeropoint, an agreement
between our photometry results and the corresponding literature values
is necessary.  When comparing MSTO color, however, one needs to
consider that different parts of the field of view are reddened by
different amounts, as we showed in our extinction map
(Fig. \ref{extmap}).

We estimated our main-sequence turnoff $V_{TO}$ magnitude to be 18.2
and the MSTO color $(V-I)_{TO}$ to be around 0.88 mag; both were 
determined by eye from the raw photometry CMD (Fig. \ref{cmd_raw}).

These values were compared to the results of Rosenberg et al. (2000)
\markcite{rosenberg00} (hereafter RB) and C\&O. RB find $V_{TO} \sim
18.2$ and $(V-I)_{TO} \sim 0.92$; both values are again determined by
eye.  C\&O report $V_{TO} \sim 18.15$ and $(B-V)_{TO} \sim 0.65$
(tabulated). Using their reddening estimate of $E_{B-V} \sim 0.22$,
the relations in Cardelli, Clayton, \& Mathis (1989), and a set of
isochrones provided by Don VandenBerg for $[Fe/H] = -1.41$ (VandenBerg
2000, private communication, based on evolutionary models by
VandenBerg et al. 2000; hereafter VDB)\markcite{vandenberg00}, the
C\&O value for $(V-I)_{TO}$ can be calculated to be approximately
0.93.

One discrepancy between our results and these literature values is
therefore that our value for $(V-I)_{TO}$ is approximately 0.05 mag
bluer than what is quoted by the authors mentioned above.  The
existence of variable reddening may be a likely explanation for this
discrepancy, especially when one considers which part of the cluster
was observed.

RB concentrate on the southern and eastern parts of the cluster for
their longer exposures and the center of NGC 3201 for their shorter
exposures.  All of these three regions are, according to our reddening
map (Fig. \ref{extmap}), regions of higher extinction with respect to
our fiducial region.

C\&O observe regions just to the north and west of the cluster center
and the cluster center itself.  While the north and west regions
suffer only low reddening (Fig. \ref{extmap}), the area within $\sim$
7 arcmin to the north and west of the cluster center is obscured by a
smaller feature not visible in the SFD map (see
Fig. \ref{schlegelmap}). At least 5 of 7 C\&O fields are very likely
affected by this dense obscuration feature.

While a quantitative (star-by-star) magnitude comparison between our
data and the data obtained by RB and C\&O is not possible, we conclude
that their estimates for the MSTO color could very well be influenced
by the fact that the regions they observed were subject to extinction
somewhat higher than the average value over our field of view.

We note that these comparisons focus on the MSTO colors; the vertical
nature of the MSTO makes it difficult to compare the $V$ mag of the
MSTO at sufficiently high precision.

%--------------------------------------------------------------------------
\subsection{Basic Isochrone Fitting and Determination of Reddening Zeropoint}
%--------------------------------------------------------------------------

We examined three different possibilities of finding the $E_{V-I}$
zeropoint to add to our extinction map.  All three methods are
outlined below. For each case, it is worth pointing out that a direct
comparison with literature values should be taken with caution since
usually, only a single numerical value for the reddening is given,
whereas our result is a reddening matrix of which every single element
needs to be added to a zeropoint offset.

%--------------------------------------------------------------------------
\subsubsection{Using Isochrone Fits}
%--------------------------------------------------------------------------

\placefigure{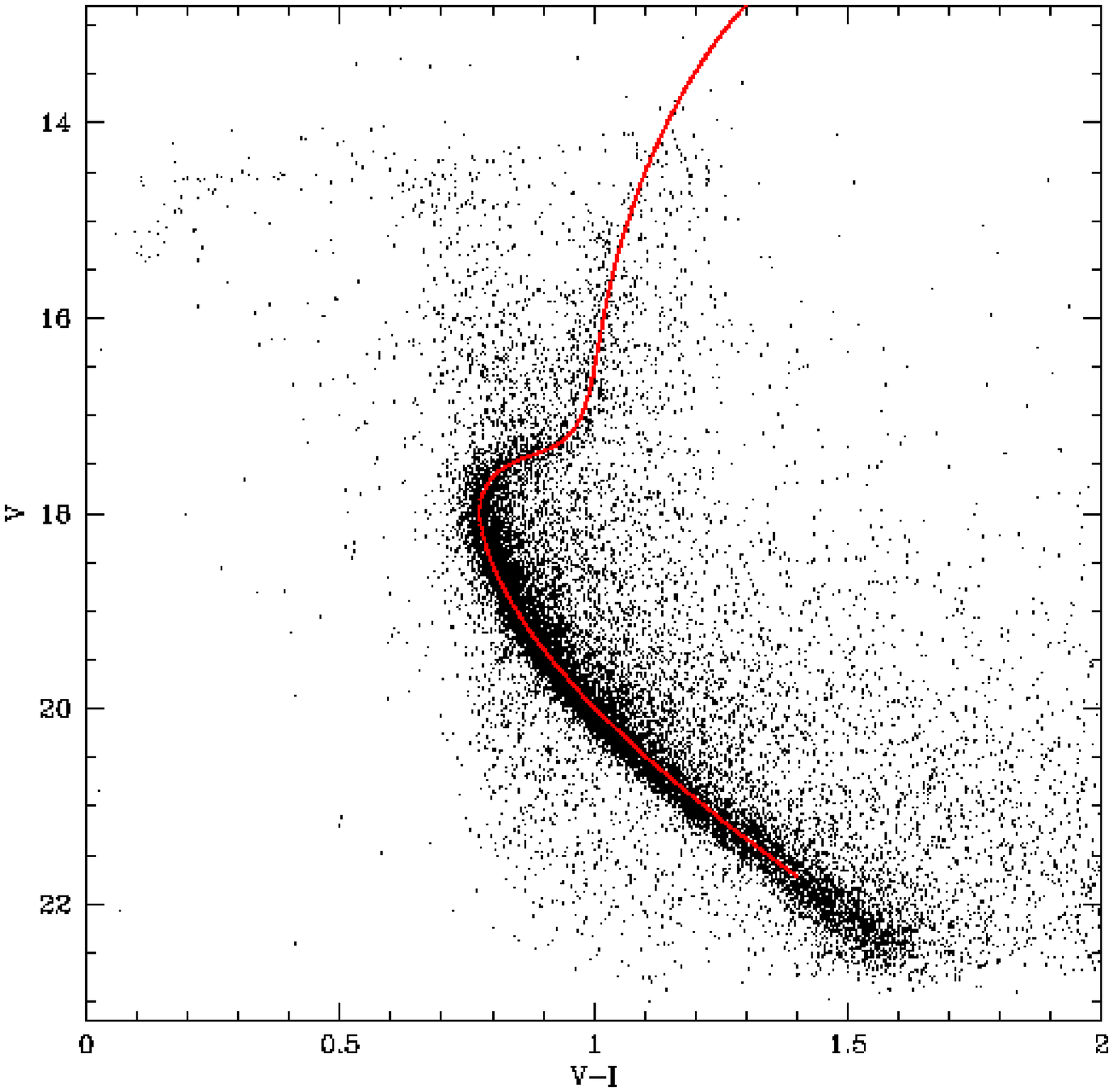}

Simultaneously fitting age, distance, and reddening to the
aforementioned set of VDB isochrones produced the fit shown in
Fig. \ref{isocmd}. It should be noted that since the point of this
publication is outlining our dereddening method as well as creating an
extinction map for NGC 3201, we only performed some basic isochrone
fitting here. For an estimate of $[Fe/H]$, we adopted the isochrone
with metallicity closest to the GW value of -1.42.

For the fit of Fig. \ref{isocmd}, $[Fe/H]=-1.41$, $d \sim 4.5$ kpc,
age = 18 Gyrs, and the $E_{V-I}$ zeropoint to be added to the values
in Fig. \ref{extmapgrid} is 0.15 mag.  The age is clearly on the high
end of things, but the 18 Gyr isochrones produced by far the best
fits, even when $[Fe/H]$ and the distance were slightly modified. The
value for the distance is slightly below the one quoted by Harris
(1996), but fairly close to the two values derived using slightly
different methods by C\&O.

When adding the mean $E_{V-I}$ of our reddening map (86 mmag) to the
zeropoint, the average $E_{V-I}$ for the cluster is approximately
0.24, which is slightly below the estimates of, e.g., Cacciari (1984)
whose $E_{V-I} \sim 0.32$, and of Harris (1996) and RB who quote the
same value.

\placefigure{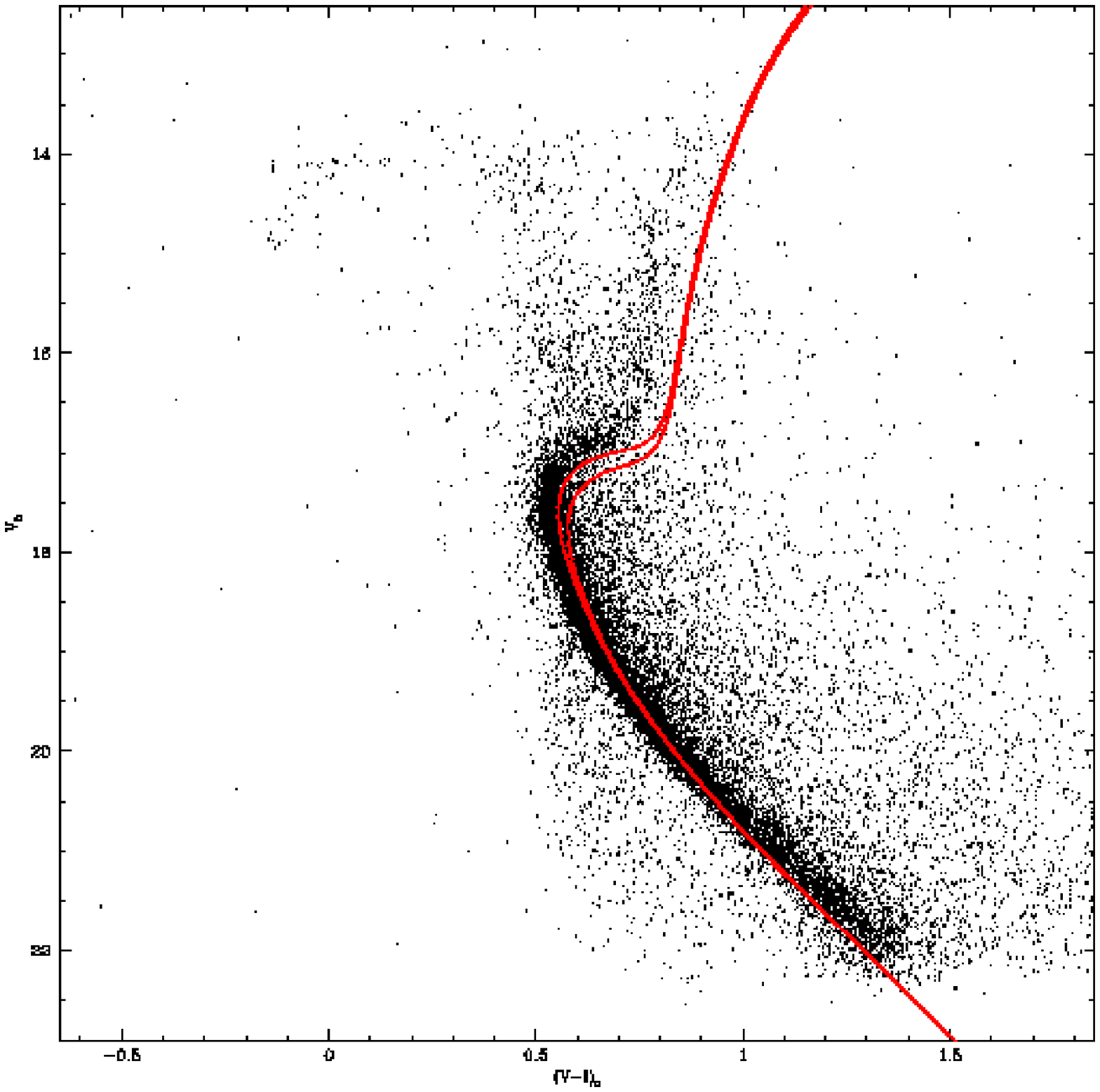}

In order to justify our choice of isochrone parameters and to
eliminate the possibility of multiple factors conspiring against us in
the attempt to find a good isochrone fit to the data, we illustrate in
Fig. \ref{badfit} what the CMD would look like if we were to strictly
adopt literature values from Harris (1996) ($[Fe/H]=-1.58$,
$E_{V-I}=0.325$, d = 5.2 kpc).  Plotted are a pair of isochrones with
respective ages 12 and 14 Gyrs on top of the data\footnote{In order to
have a valid basis for comparison, we needed to apply an $E_{V-I}$
zeropoint of $0.325-0.086 = 0.239$ to the differentially dereddened
data, where 0.325 is the Harris (1996) $E_{V-I}$ value and 0.086 is
the mean $E_{V-I}$ value of our extinction map.}.  The agreement
between data and fit is not as good as in Fig. \ref{isocmd}.

%--------------------------------------------------------------------------
\subsubsection{Using SFD Maps}
%--------------------------------------------------------------------------

As stated in Section 3.1, the mean $E_{V-I}$ of the difference map
between the SFD and our data is $401\pm46$ mmag.  The average
$E_{V-I}$ value for this difference map corresponds to the $E_{V-I}$
zeropoint to add to our data to obtain absolute reddening values.

Calculating the total reddening in this fashion gives $E_{V-I} \sim
0.49$ which is significantly higher than the literature
values. Furthermore, using the isochrones mentioned in the previous
subsection, we were not able to produce a fit to the data due to the
fact that $E_{V-I}$ was too high.

%--------------------------------------------------------------------------
\subsubsection{Using RRLyrae Stars}
%--------------------------------------------------------------------------

\placefigure{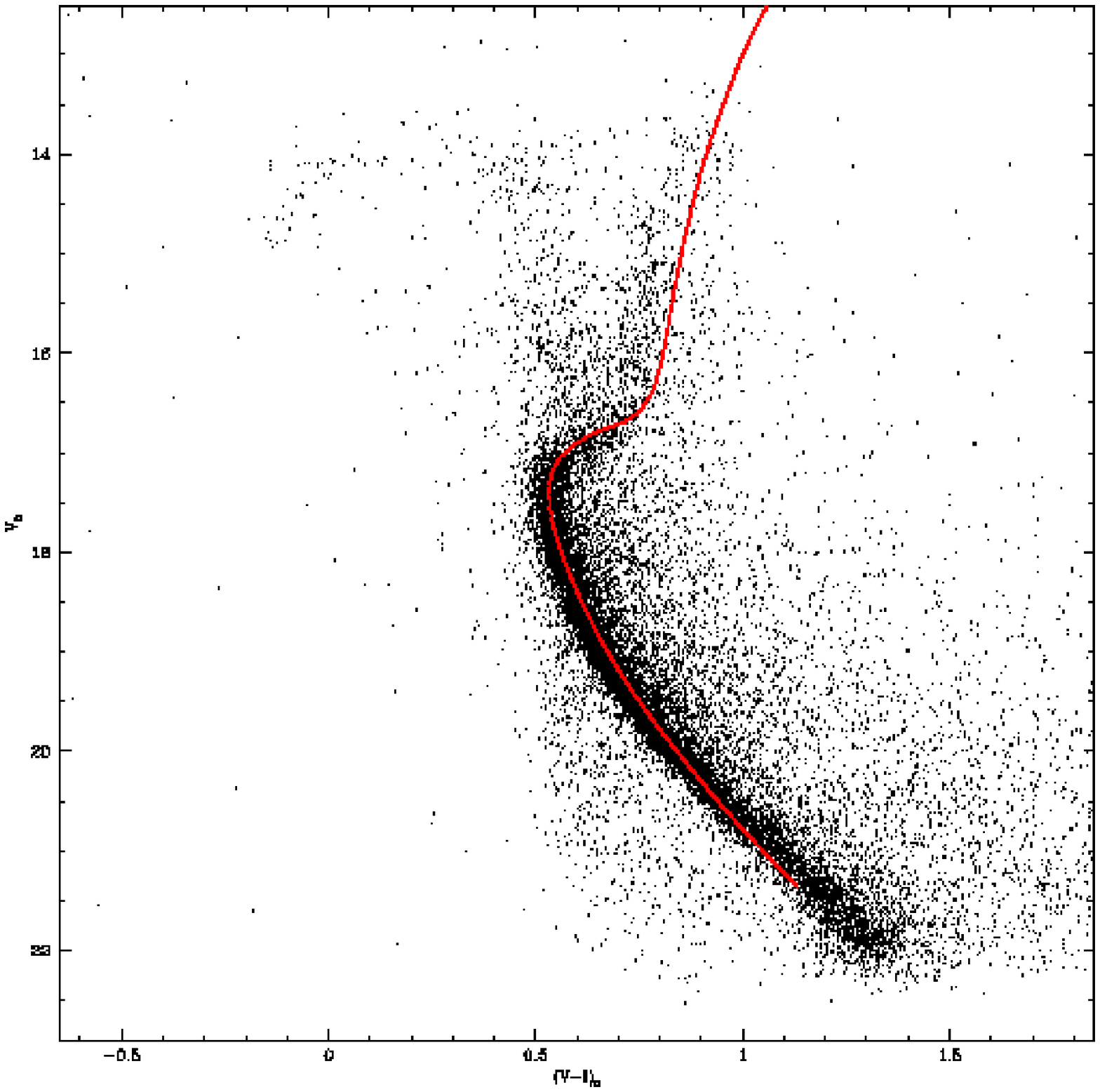}

Another method of attempting to find a reddening zeropoint is to
simply adopt one from the literature and model the thus dereddened
data with a set of isochrones. Here, we took Cacciari's (1984) value
of $E_{V-I} \sim 0.32$ which she calculated using RRLyrae stars. Since
this corresponds to the average reddening for the whole cluster, we
had to subtract the mean $E_{V-I}$ of her RRLyrae stars (in our field
of view) with respect to our fiducial region. This value turned out to
be 79 mmag.

We failed, however, to find a good isochrone fit to the thus
dereddened data, as is shown in Fig. \ref{rrlyraecmd}. In order to get
the MSTO points of the data and the isochrone to agree, we had to
lower $[Fe/H]$ to -2.01, a value quite a bit below literature values
such as -1.41 (GW) or -1.58 (Harris 1996).  The distance modulus here
is 13.4, and the age 14 Gyrs.  As one can easily see, the fit fails to
hit the subgiant branch due to the ``young'' isochrone age of 14
Gyrs. Older age isochrones could not be shifted far enough toward the
dereddened MSTO location to produce a fit, even when the metallicity
was lowered to the lowest available value of -2.31.

%--------------------------------------------------------------------------
\subsubsection{Comments on the Adopted Reddening Zeropoint}
%--------------------------------------------------------------------------

Given that we were not able to produce a fit to the data using the VDB
isochrones when we calculated the reddening zeropoint using either the
SFD maps or the Cacciari (1984) RRLyrae stars, we conclude that our
best estimate for the $E_{V-I}$ zeropoint is 0.15 mag, giving an
average $E_{V-I} \sim 0.24$ for the cluster as a whole. This value
falls approximately $1.5 \sigma$ below the estimate by Cacciari (1984)
and the value tabulated by Harris (1996).

We could force our reddening zeropoint to agree with these past
results by applying a systematic shift to our $I$ magnitudes of 0.08
to brighter values which would give a distance to NGC 3201 of 4.8 kpc.
We have, however, no reason to adopt such a shift. The transformations
to standard magnitudes as well as the application of the aperture
corrections (see Section 2.1) underwent repeated thorough examinations
and appear to be correct. Since no other indication throughout our
data reduction and analysis suggests a systematic error in the $I$
band magnitudes, either, we simply cannot justify adjusting our $I$
magnitudes in order to compensate for the $1.5 \sigma$ effect
described above.

%--------------------------------------------------------------------------
\section{Summary and Concluding Remarks}
%--------------------------------------------------------------------------

\begin{figure}
\plotone{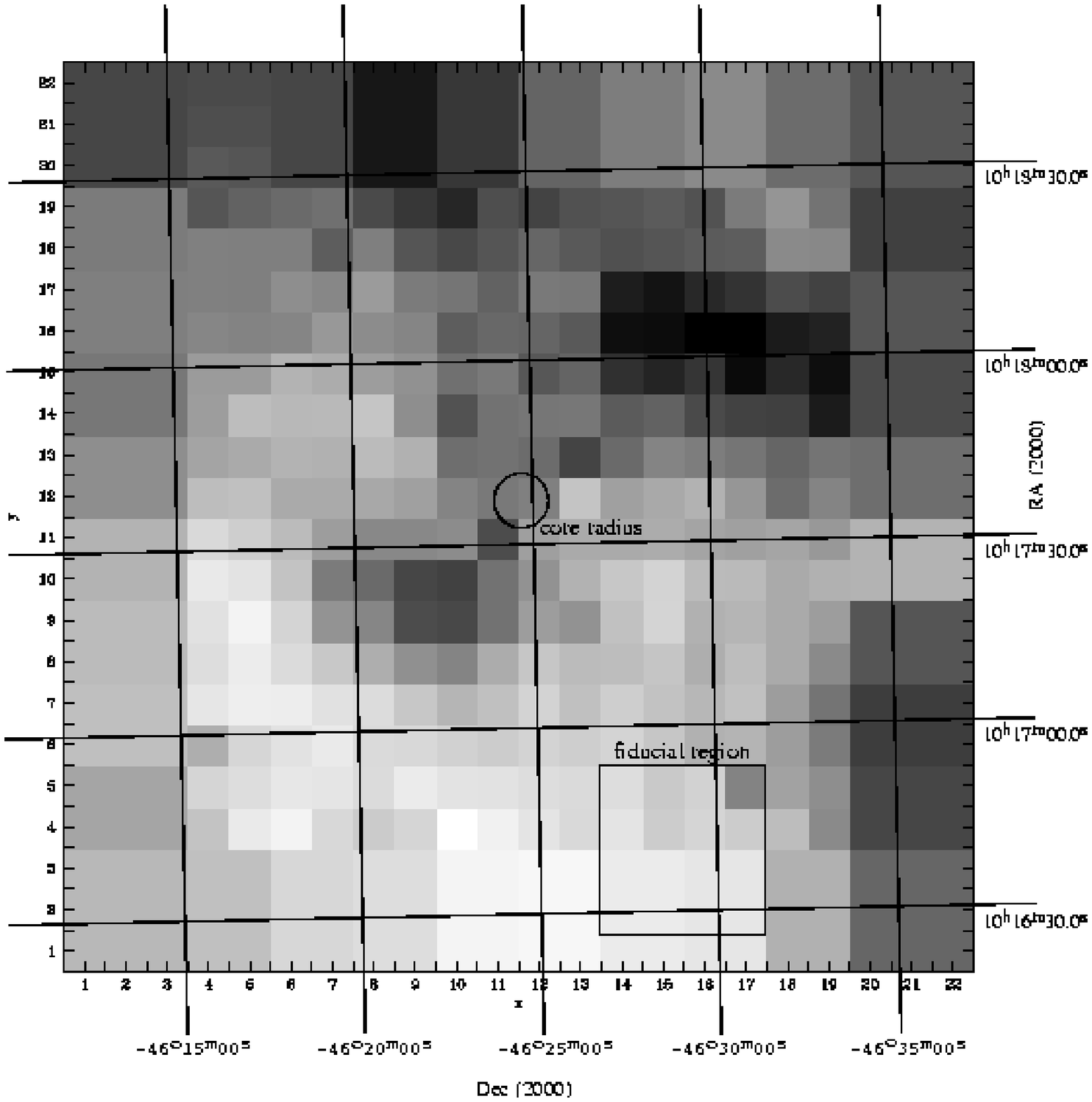}
\caption{The extinction map. East is up, north is to the left. The
darker the color of a subregion, the more extinction is occurring in
it (see Fig. \ref{extmapgrid}).  Note the ``ring'' of larger
subregions (with lower density of stars) around the center of the
field of view. The size of every small subregion is approximately 70
arcsec on a side. For reference, we included an illustration of the
core radius of NGC 3201 (from Harris 1996) around the location of the
cluster center on the extinction map as well as the location of the
fiducial region described in Section 2.2. Furthermore, we overlaid a
grid for coordinate reference. The coordinate axes $x$ and $y$
correspond to the ones used in Fig. \ref{extmapgrid}.  The average
reddening with respect to the extinction occurring in the fiducial
region is $E_{V-I}=86 \pm 61$ mmag.
\label{extmap}}
\end{figure}

\begin{figure}
\plotone{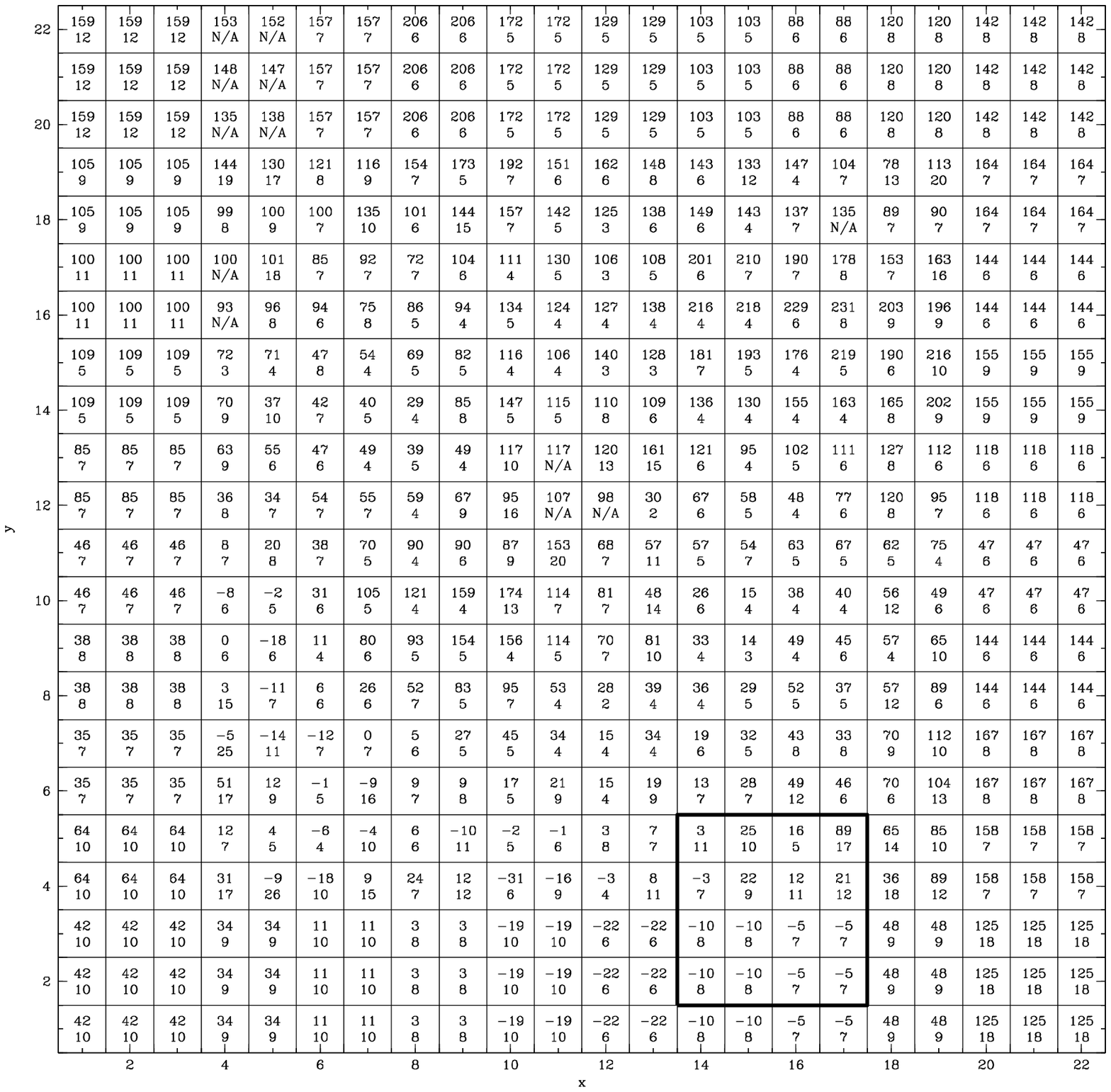}
\caption{The $E_{V-I}$ values with respect to the fiducial region for
the individual pixels of our reddening map in millimagnitudes.  The
pixels correspond to the subregions in Fig. \ref{extmap} with the
respective values for $x$ and $y$ coordinates.  The top number in
every pixel is the value for $E_{V-I}$ obtained with the method
outlined in Section 2.2. The bottom number is the associated error, as
described in Section 2.2, item 7. The pixels where the error value is
N/A are the ones with an insufficient number of stars, so the
reddening results were obtained by interpolation from neighboring
values (see Section 2.2, item 6).  For reference to figures
\ref{extmap}, \ref{ngc3201_final}, \ref{schlegelmap}, and
\ref{data-map} (same orientation), the location of the fiducial region
is indicated in the lower right part of the figure.  The average
reddening with respect to the mean reddening occurring in the fiducial
region is $E_{V-I}=86 \pm 61$ mmag.  In order to obtain a map of
absolute reddening values, an $E_{V-I}$ zeropoint has to be added to
the data. We discuss the various zeropoints we examined in Section
4.1. Finally, to obtain an absolute value for $E_{B-V}$, one needs to
multiply the absolute (i.e., including the zeropoint) $E_{V-I}$ value
by 0.6468 (Cardelli, Clayton, \& Mathis 1989).
\label{extmapgrid}}
\end{figure}

\begin{figure}
\plotone{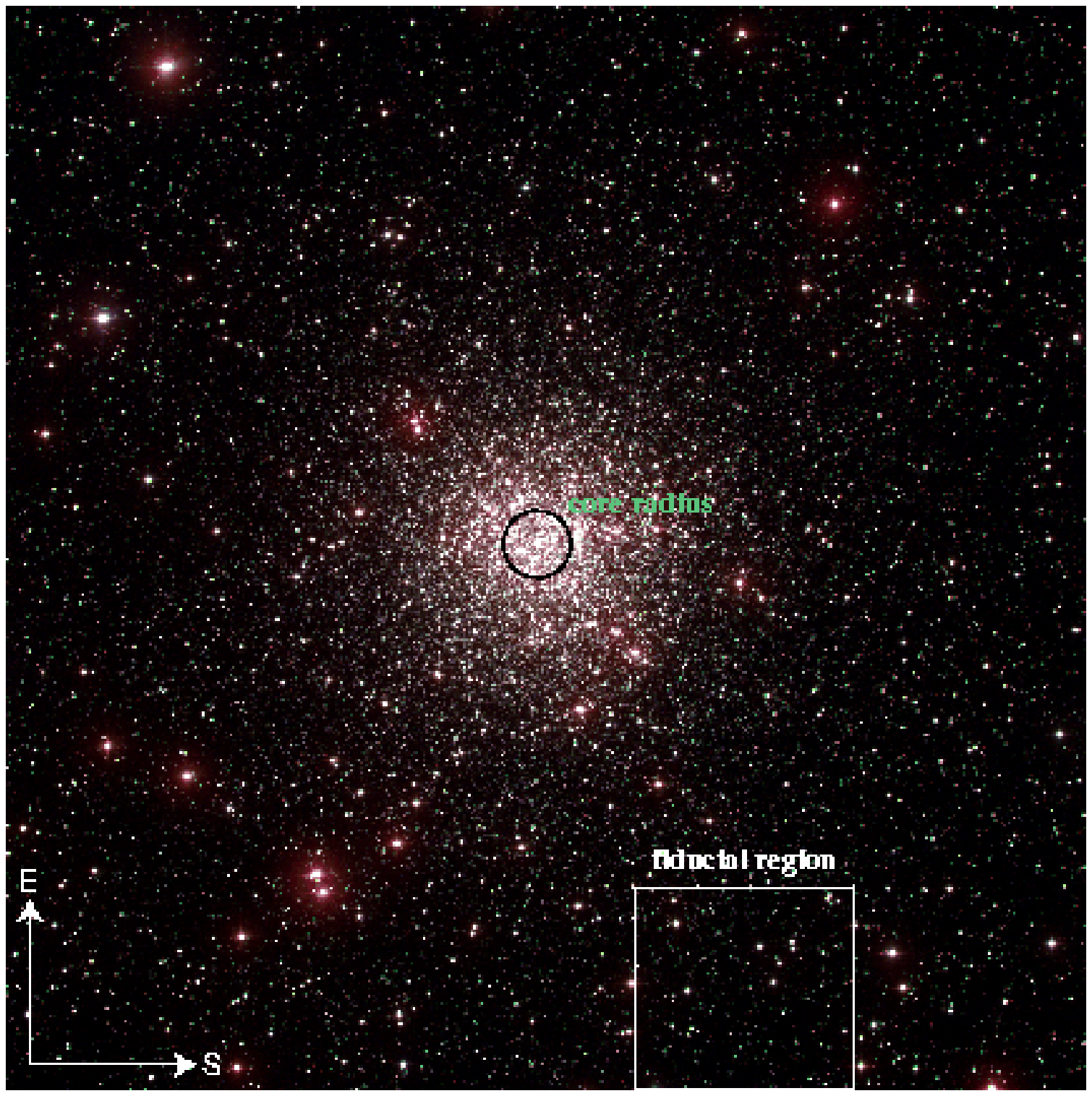}
\caption{An image of NGC 3201 with illustration of core radius and
location of fiducial region for purposes of comparison with Figures
\ref{extmap}, \ref{extmapgrid}, \ref{schlegelmap}, and
\ref{data-map}. The size of the field of view shown in this Figure is
about 81\% (area) that of the field of view of the Figures mentioned
above.
\label{ngc3201_final}}
\end{figure}

\begin{figure}
\plotone{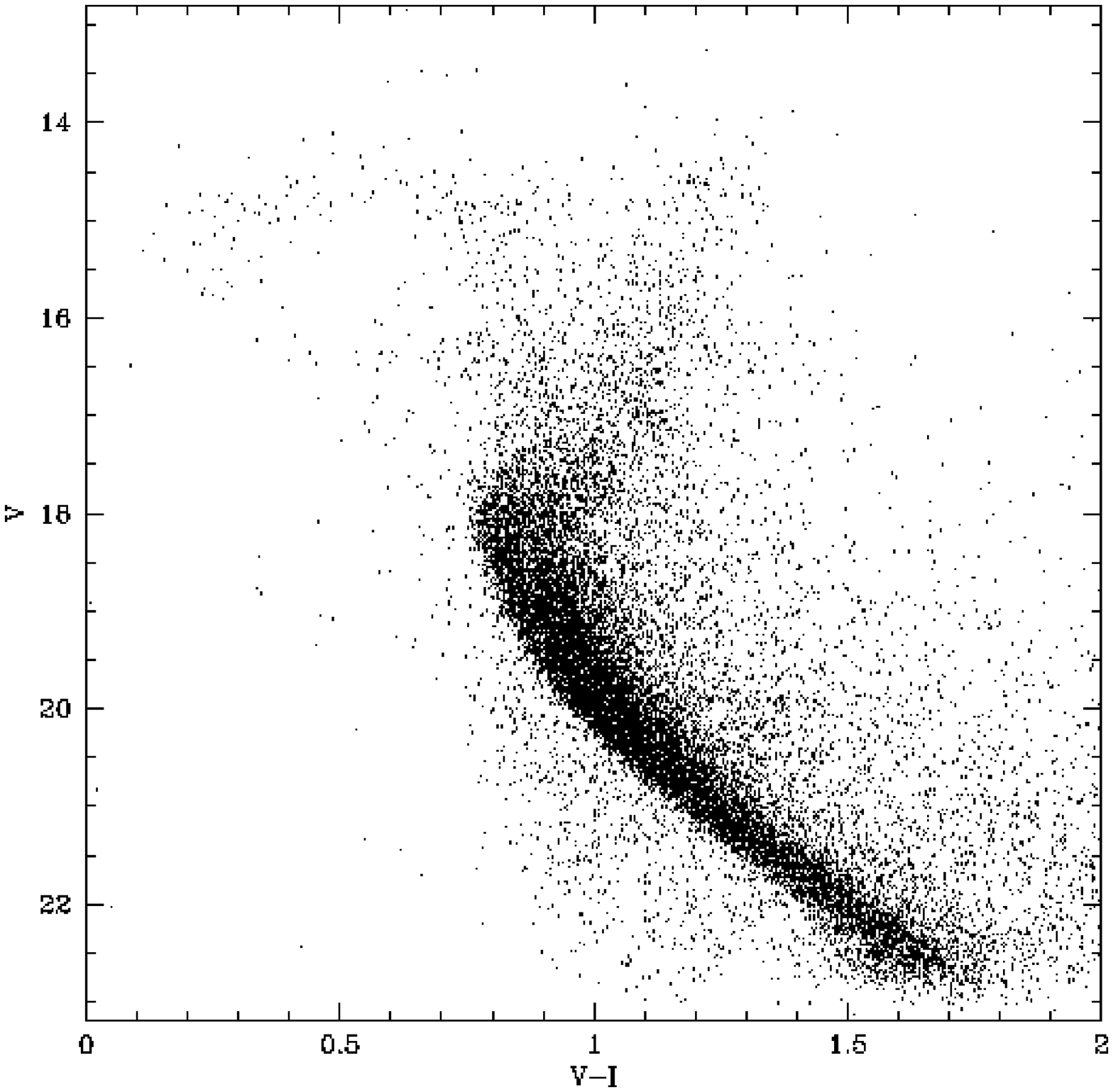}
\caption{The raw CMD of NGC 3201 before applying any dereddening.
\label{cmd_raw}}
\end{figure}

\begin{figure}
\plotone{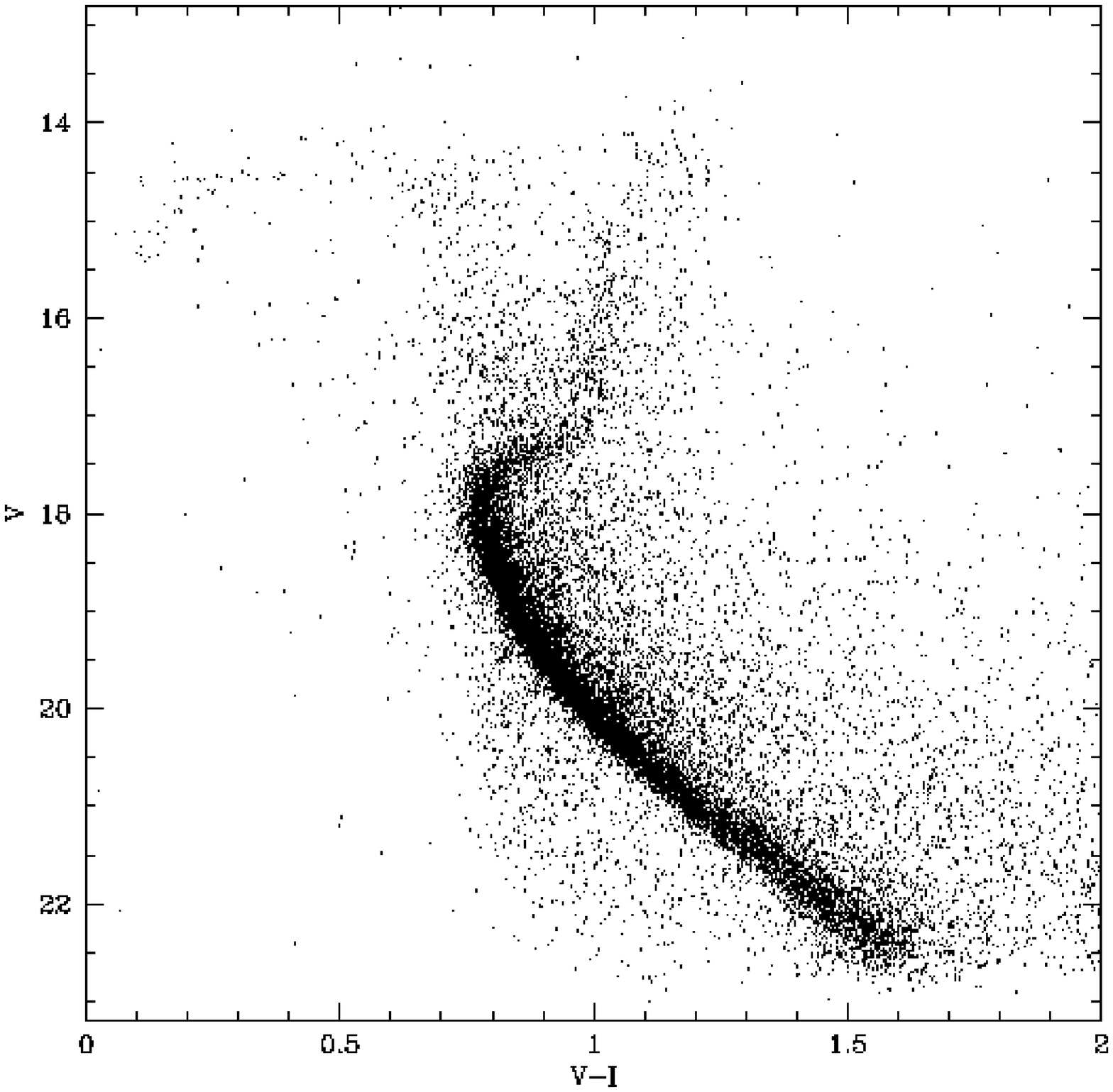}
\caption{The internally dereddened CMD of NGC 3201.  $V$ and $V-I$
indicate the location of the datapoints after the differential
reddening with respect to the fiducial region was corrected for, i.e.,
after the inverse values of Fig. \ref{extmapgrid} were applied to the
data. No reddening zeropoint is applied to the data in this plot.
\label{cmd_final}}
\end{figure}

\begin{figure}
\plotone{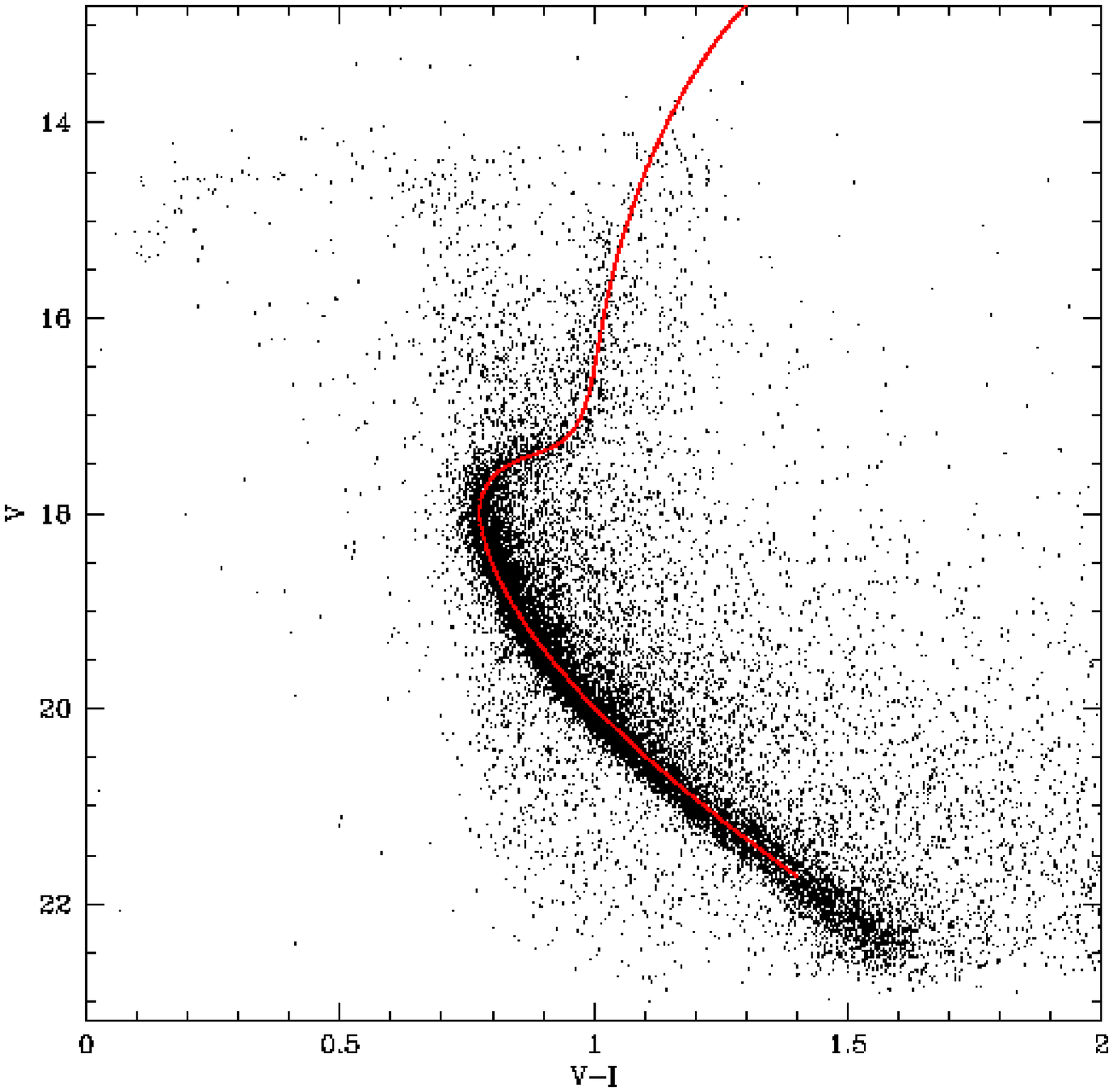}
\caption{The best fit to our data produced by the VDB isochrones. For
this fit, $[Fe/H]=-1.41$, $d \sim 4.5$ kpc, age = 18 Gyrs, and the
reddening zeropoint $E_{V-I} \sim 0.15$ to be added to the values in
Fig. \ref{extmapgrid}.  The average $E_{V-I}$ for the cluster is
approximately 0.24 in this plot.
\label{isocmd}}
\end{figure}

\begin{figure}
\plotone{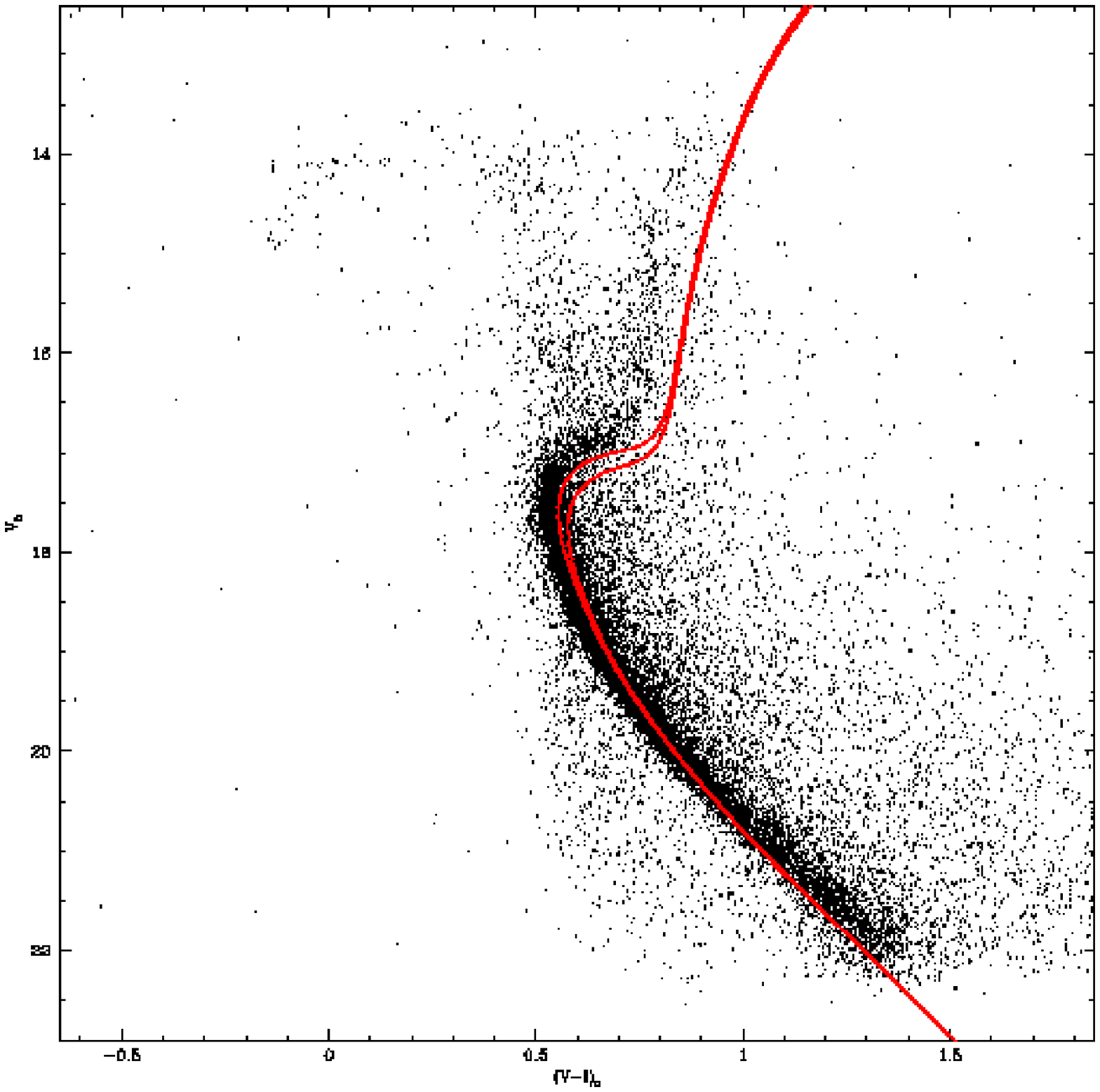}
\caption{The CMD of NGC 3201, dereddened to the value of $E_{V-I} =
0.325$ (Harris 1996), plus the isochrones for ages of 12 (top) and 14
(bottom) Gyrs with values of $[Fe/H]=-1.54$ (closest value to Harris'
-1.58) and d = 5.2 kpc.  The agreement between data and fit is better
in Fig. \ref{isocmd}.
\label{badfit}}
\end{figure}

\begin{figure}
\plotone{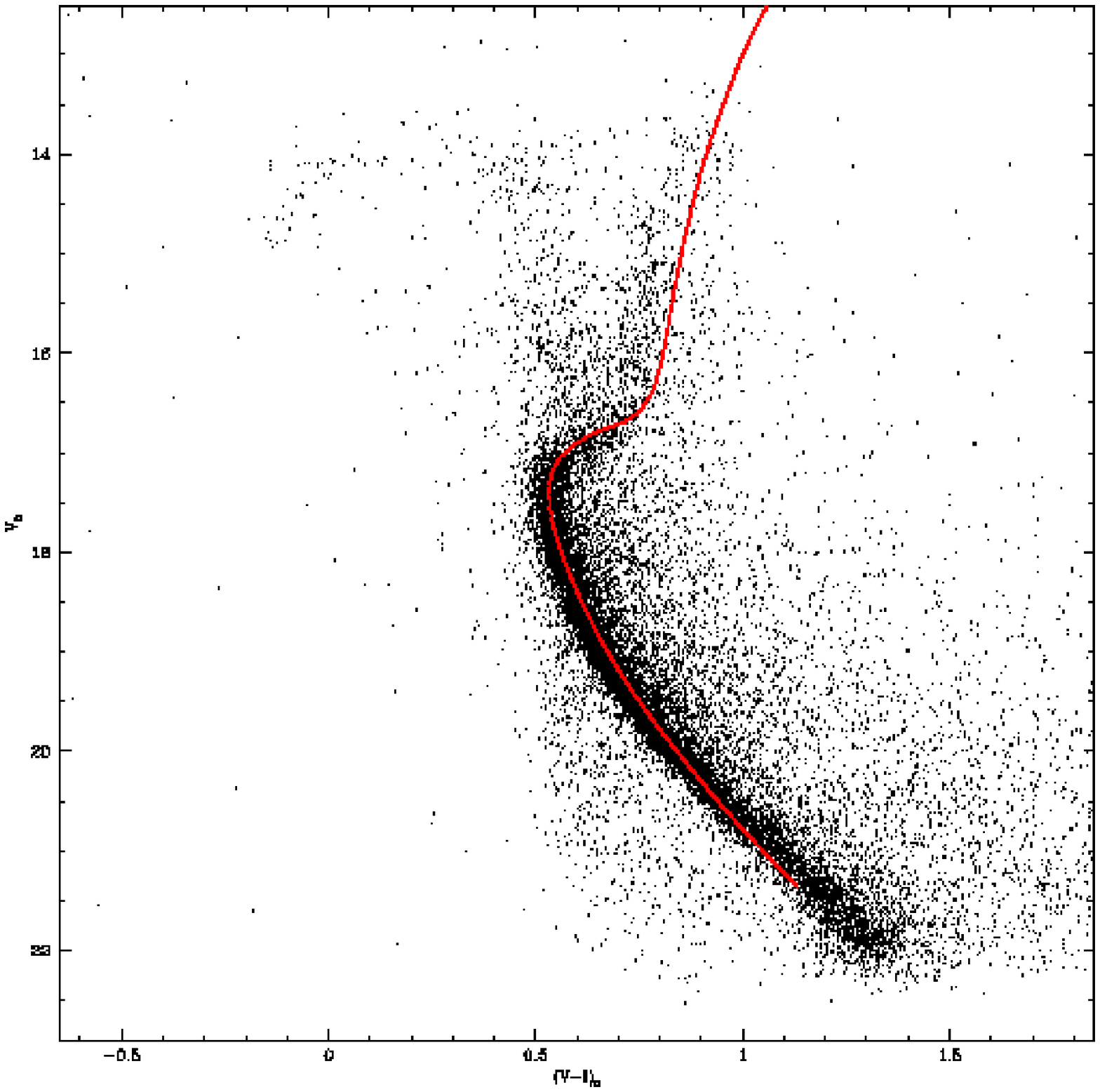}
\caption{The fit to the data dereddened by adopting the $E_{V-I}$
zeropoint of 0.24, using Cacciari's (1984) RRLyraes, and applying our
reddening map.  $[Fe/H]=-2.01$ (much lower than literature values),
age = 14 Gyrs, and distance modulus = 13.4. As one can see, the fit
fails to reproduce the loci of the datapoints. We therefore conclude
that the reddening zeropoint of Fig. \ref{isocmd} is correct.
\label{rrlyraecmd}}
\end{figure}

In the process of finding eclipsing binary stars in NGC 3201, we
noticed the existence of variable reddening of up to 0.2 mag in
$E_{V-I}$ on a scale of arcminutes. Using our internal dereddening
method outlined in Section 2.2, we created an extinction map which is
shown in Figures \ref{extmap} and \ref{extmapgrid}. Applying the map
to our raw data (Fig. \ref{cmd_raw}) significantly improved the
appearance of the CMD (Fig. \ref{cmd_final}).

\begin{figure}
\plotone{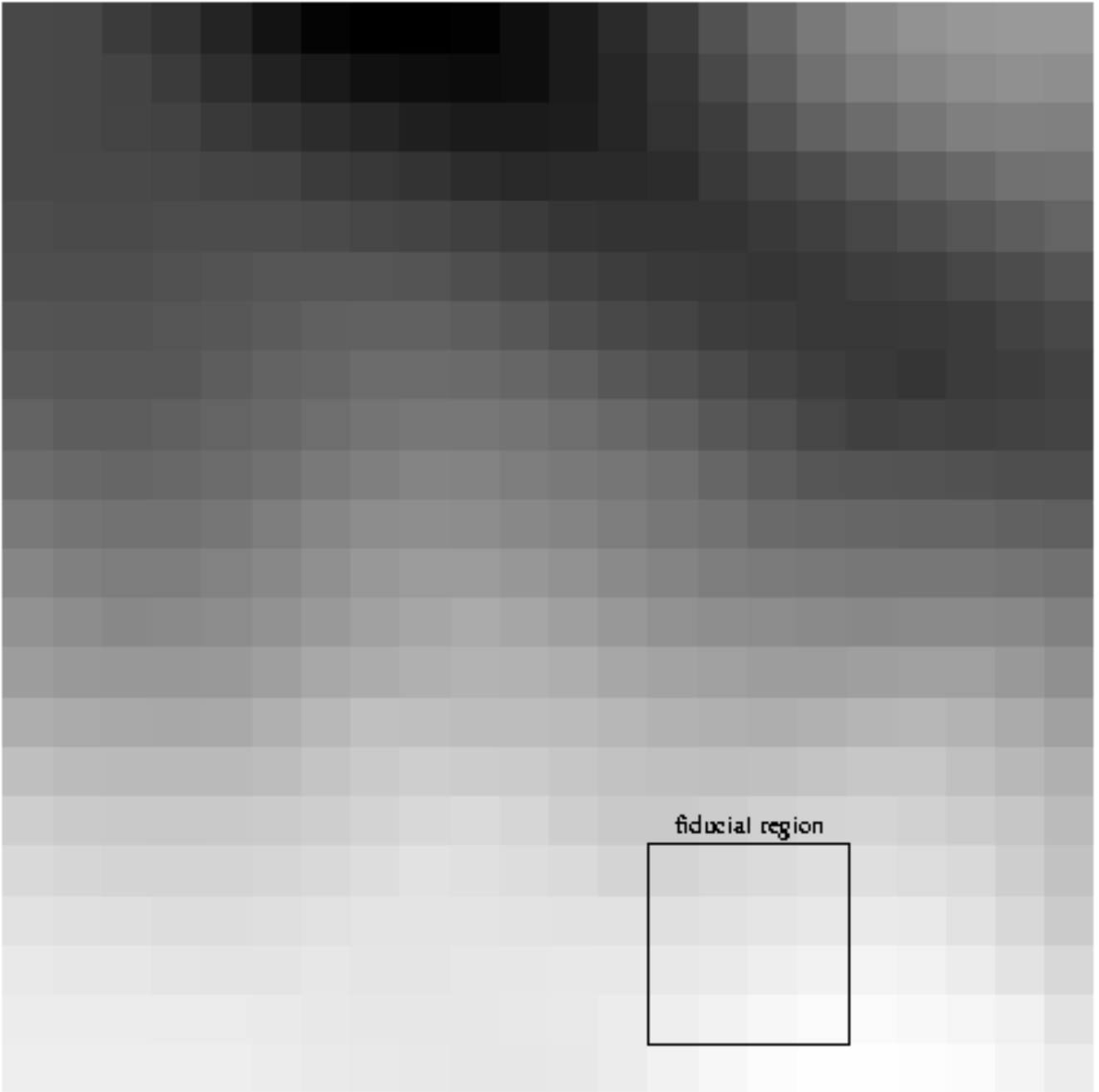}
\caption{A graphical representation of the SFD data in the region of
NGC 3201 that we observed. The location of our fiducial region is
given for reference; the orientation of the map is the same as in
Figures \ref{extmap}, \ref{extmapgrid}, \ref{ngc3201_final}, and
\ref{data-map}. Common features are easily recognizable such as the
ridge of obscuring material extending from the top left part of the
map to the right center. A number of smaller-scale features, however,
do not show up on the SFD map (see Fig. \ref{data-map}).  The average
reddening is $E_{V-I}=487 \pm 59$ mmag.
\label{schlegelmap}}
\end{figure}

\begin{figure}
\plotone{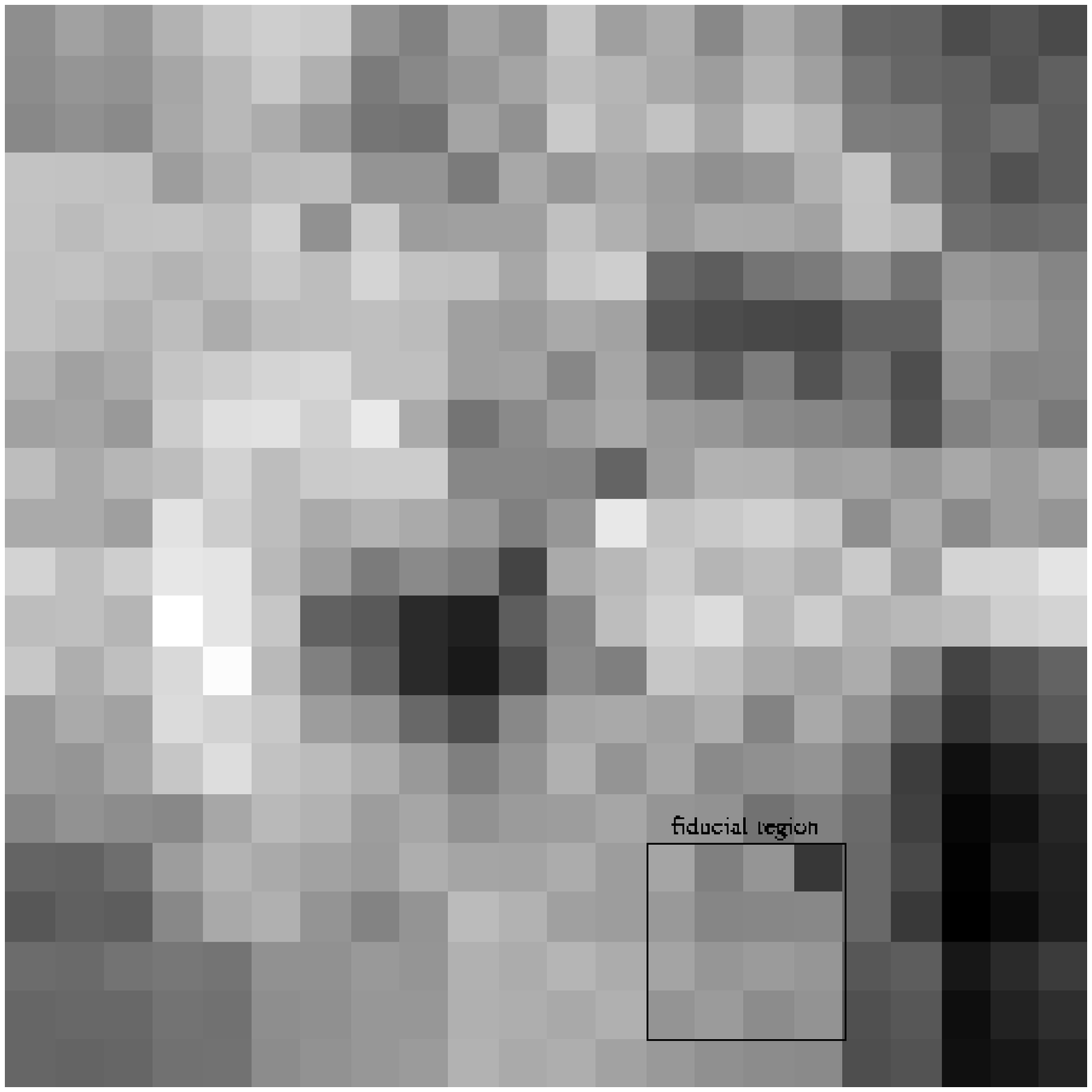}
\caption{A difference plot between our extinction map
(Fig. \ref{extmap}) and the SFD map (Fig. \ref{schlegelmap}) with the
same orientation. Darker regions correspond to areas where our map
indicates more obscuration relative to the average value than the SFD
data; lighter regions are places where the two maps agree very well.
The average reddening is $E_{V-I}=401 \pm 46$ mmag, which corresponds
to the SFD-map-zeropoint (see Section 4.2.2). Note that the standard
deviation is lower than the one of the SFD map and the one of our
extinction map.
\label{data-map}}
\end{figure}

Comparison between our extinction map and the SFD map of the same
region (Fig. \ref{schlegelmap}) showed that the same larger-scale
features exist in both maps. Our map displays some additional,
smaller-scale features which are absent in the SFD maps (see
Fig. \ref{data-map}).

The $E_{V-I}$ zeropoint which needs to be added to the numbers in
Fig. \ref{extmapgrid} in order to get absolute $E_{V-I}$ values is
0.15. This value is below literature results (Cacciari 1984, Harris
1996) by approximately $1.5 \sigma$, but produced by far the best VDB
isochrone fit to the data.  The zeropoint determined with the help of
the SFD maps gives $E_{V-I} \sim 0.49$ as the average value across NGC
3201 which is higher than literature values which supports the
statement by Arce \& Goodman (1999) that the SFD maps overestimate the
reddening in regions of high extinction.

The results from this work will be essential in our binary star
research where high-quality photometry of every binary system is
necessary for distance determinations. A vital condition to obtaining
these measurements is of course the knowledge of the exact extinction
the star under investigation is suffering.

Furthermore, studies like this may be useful in determining properties
of the ISM itself such as examining a possible dependence of $R_V$
upon position in the field of view or giving insight into the
distribution and properties of the dust along the line of sight.

%--------------------------------------------------------------------------
\acknowledgments
%--------------------------------------------------------------------------

This research was funded in part by NSF grants AST96-19632 and
AST98-20608.  We would like to sincerely thank Alex Athey, Kristin
Chiboucas, Robbie Dohm-Palmer, Ivan King, and Don VandenBerg, as well
as the anonymous referee, for their help, advice, and pointing out
errors during the various stages of the creation of this publication.

%--------------------------------------------------------------------------
% References
%--------------------------------------------------------------------------

%\pagebreak

%--------------------------------------------------------------------------
% Table
%--------------------------------------------------------------------------

%\newpage
%
%\begin{figure}
%\plotone{table1.ps}
%\end{figure}

%--------------------------------------------------------------------------
% Figures and Captions:
%--------------------------------------------------------------------------

\clearpage

\newpage

\newpage

\newpage

\newpage

\newpage

\newpage

\newpage

\newpage

\newpage

\newpage

\newpage

\end{document}